\documentclass{article}

\usepackage{PRIMEarxiv}

\usepackage[utf8]{inputenc} % allow utf-8 input
\usepackage[T1]{fontenc}    % use 8-bit T1 fonts
\usepackage{hyperref}       % hyperlinks
\usepackage{url}            % simple URL typesetting
\usepackage{booktabs}       % professional-quality tables
\usepackage{amsfonts}       % blackboard math symbols
\usepackage{nicefrac}       % compact symbols for 1/2, etc.
\usepackage{microtype}      % microtypography
\usepackage{lipsum}
\usepackage{fancyhdr}       % header
\usepackage{graphicx}       % graphics
\graphicspath{{media/}}     % organize your images and other figures under media/ folder

\usepackage{setspace}
\usepackage{booktabs}
\usepackage{amsfonts}
\usepackage{amsmath}
\usepackage{commath}
\renewcommand{\Omega}{\varOmega}

\newcommand{\DU}{\mathrm{D}}

\newcommand{\grad}{\boldsymbol{\nabla}}

\newcommand{\crsc}[1]{\left[ #1 \right]}

\newcommand{\veb}[1]{\mathbf{#1}}

\newcommand{\rnl}{\mathrm{Re}}

\newcommand{\etal}{\textit{et al.\ }}

\title{Development of Smoothed Particle Hydrodynamics Method for Modeling Active Nematics
}

\author{
  Roozbeh Saghatchi, Deniz Can Kolukisa, and Mehmet Yildiz* \\
  1- Faculty of Engineering and Natural Sciences, Sabanci University, Tuzla, 34956 Istanbul, Turkey\\
		2- Integrated Manufacturing Technology Research \& Application Center, Sabanci University, Tuzla, 34956 Istanbul, Turkey\\
		3- Composite Technologies Center of Excellence, Sabanci University-Kordsa, Pendik, 34906 Istanbul, Turkey\\
  \texttt{*Corresponding author; mehmet.yildiz@sabanciuniv.edu} \\
}

\begin{document}
\maketitle

\begin{abstract}
This paper proposes a novel GPU-based active nematic flow solver based on the smoothed particle hydrodynamics (SPH) method. Nematohydrodynamics equations are discretized using the SPH algorithm, and the periodic domain is enforced using the periodic ghost boundary condition. Flow behavior, nematic ordering, topological defects, vorticity correlation is calculated and discussed in detail. Due to the high particle resolution, the spectrum of the kinetic energy with respect to the wavenumber is calculated, and its slope a the different length scales discussed. To exploit the SPH capabilities, pathlines of nematic particles are evaluated during the simulation. Finally, the mixing behavior of the active nematics is calculated as well and described qualitatively. The effects of two important parameters, namely, activity and elastic constant are investigated. It is shown that the activity intensifies the chaotic nature of the active nematic by increasing the pathline and mixing efficiency, while the elastic constant behaves oppositely.
\end{abstract}

% keywords can be removed

	\section{\label{sec:int}Introduction}
Active matters have generated great interest among researchers due to their pivotal importance in a vast variety of applications which mainly arise from their out of thermodynamic equilibrium characteristics because of the existence of living entities that consume energy and convert it into some forms of mechanical energy. Some examples include, but not limited to either natural systems such as fish, birds, and animal herds, or synthetic systems composed of active energy-consuming components which convert energy from light or chemical gradients to perform work \cite{synthetic}.

\begin{figure}[hbt!]
	\centering
	\includegraphics[width=2.0in]{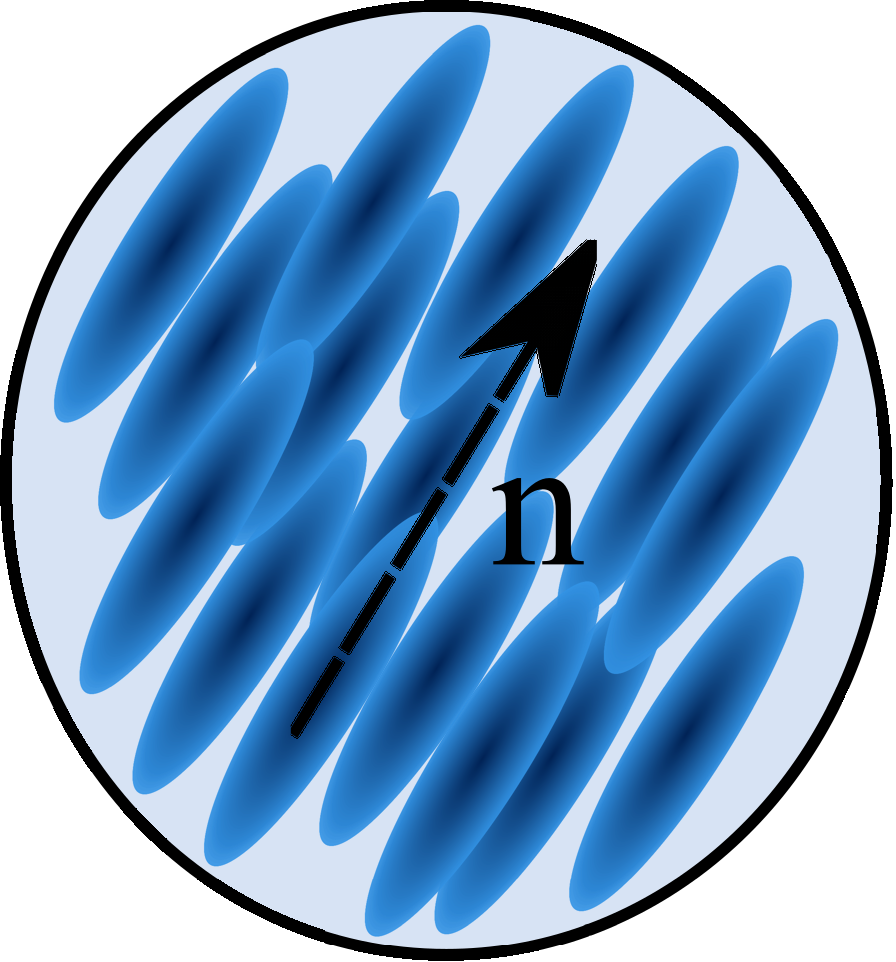}
	\caption{Schematic representation of active nematics.}
	\label{fig:schematics}
\end{figure}

Active nematics are special types of active systems that consist of rod-like nematic liquid crystals with head-tail symmetry (\figref{fig:schematics}) \cite{amin2018}. Active nematics form an orientationally ordered but not positionally ordered state. The destruction of long-ranged order initiates from the existence of spontaneous topological defects due to the activity of nematics which leads to turbulent-like behavior, namely active turbulence \cite{amin2017turb}. Microtubule/Motor-Protein mixtures \cite{Microtubule}, bacterial suspensions \cite{bacterial}, and cell layers \cite{amin2017cell} are some of the active nematics examples.

Many experimental, theoretical and numerical studies have been conducted to understand the different characteristics of active nematics. In this study, since our focus is on the development of a novel particle based solver for numerically modeling the complex flow characteristics of active nematics, for the in-depth theoretical and physical background as to the active nematics, interested readers are referred to \cite{amin2018}.

Based on the continuum theory, different approaches have been developed to simulate the active fluids. Henricus \etal introduced the mesoscale approach which combined the nematodynamics with the Navier-stokes equation that leads to a single equation, namely Toner–Tu equation, supplemented with a Swift–Hohenberg-type fourth-order term \cite{Wensink2012}. Although this method is simple and straightforward, it carries no information about orientation field and ordering magnitude and it exhibits the hydrodynamics fields such as active turbulence. Other approaches are Ericksen–Leslie–Parodi (ELP) method which determines the coupling between the nematic orientation field, $\veb{n}$ and the velocity field, $\veb{u}$ \cite{ericksen-leslie-parodi}, and Beris–Edwards (BE) model which introduces order parameter tensor, ${\veb{Q}}$ \cite{amin2018}. While the ELP approach is simple and gives the alignment orientation, it lacks the necessary information about the ordering magnitude. This characteristic is important when dealing with the topological defects in the active nematics \cite{amin2018}. Therefore, BE model is used in our simulation.

BE model has been utilized by many researchers who have tried to implement the BE model into various numerical methods such as Lattice Boltzmann method (LBM) and Finite Difference (FD) method. Even though LBM has been used extensively to simulate the active crystal liquids \cite{lbm2000,lbm2001}, due to its high computational cost, especially in 3D domains, hybrid LBM-FD was introduced \cite{hyblbm2004,hyblbm2007,hyblbm2011}. In this method, FD method is used to solve the nematodynamics while the LBM is utilized for the hydrodynamic part. This method has been vastly used in many investigations dealing with different aspects of active fluids such as epithelium modeling \cite{amin2017cell}, controlling dynamics and transport of liquid crystals \cite{control2021}, confinement effect \cite{confinement2021}, among others. Two of the pioneering studies based on the hybrid LBM-FD method were carried out by Thampi \etal \cite{nematiccorrelation1,nematiccorrelation2} in which they investigated the defects and correlations in active nematics. They validated the nematohydrodynamics model with the experimental studies considering the velocity and vorticity correlations. Additionally, they investigated the variation of $\pm 1/2$ defects by changing the nematic parameters. They also investigated the effect of involved parameters including activity, elastic constant, and rotational diffusion constant, on the order parameter and flow fields. According to their investigation, vorticity strongly depends on the defect in nematics. They discuss the length scale with the help of correlations of both the order parameter field and the vorticity field.

LBM method requires Eulerian lattices as grid structures and its underlying particle approach can be deemed as semi-Lagrangian. The main motivation and triggering point of the current study is the immense Lagrangian nature of the active fluids and active nematics problems, which can be capitalized by utilizing a fully Lagrangian particle-based method such as Smoothed Particle Hydrodynamics (SPH). One of the biggest advantages of the SPH method is that the convective terms in the conservation of mass and linear momentum as well as nematodynamics equations can be discretized directly with the material derivative terms thereby reducing the numerical complexity associated with the discretization of relevant nonlinear terms. Moreover, complex and deformable domains, mixtures, multiphase interfaces can also be tracked easily without using an Eulerian mesh. Lagrangian characteristics of the SPH method may be used to follow a specific nematic particle and investigate its behavior during the simulation. Besides, the mixing of the nematic particles can be illustrated easily without any additional effort.

Many challenging fluid and solid mechanics related problems have been successfully handled with the SPH methods, which for instance include free surface flows \cite{Saghatchi2014,roozbehMPC1,roozbehMPC2}, nano- \cite{Saghatchi2015} and bio-heat transfer \cite{Ghazanfarian}, multi-phase flow \cite{sphmultiphase},  electrohydrodynamics (EHD),  \cite{Shadloo2013,roozbehconfined,roozbehde}, amongst others. However, to the best of our knowledge, an attempt for simulating active nematics with the SPH method has not been realized to date. Therefore, in this study, a novel nematohydrodynamics solver based on the SPH method is developed for the simulation of the flow containing active nematics and studying the effects of important parameters involved.

%One of the challenges in SPH method is its extreme computational time due to large particle interactions. Moreover, in active nematics physiscs, there is another concern due to the existence of additional equation, namely nematodynamics, which should be calculated over time. To allay concerns about computational time, operation may be done using GPU. Earlier works have been used OpenGL to program the GPU \cite{sphcudafirst}. After introduction of CUDA, due to its advantages over OpenGL, it has been widely used by SPH researchers \cite{sphcudafirst,sphcuda}.

Utilization of high-performance parallel computational algorithms is becoming a necessity in the solution of complex physical problems in the fields of computational fluid and solid mechanics. This especially becomes inevitable for particle methods such as SPH which includes quite many neighbor interactions for each single interpolation point in the computational domain. Therefore, several studies have emerged in the last two decades, which employ multi-core/multi-thread parallelization on Central Processing Units (CPU) \cite{Ganzenmuller2007, Ihmsen2011, Marrone2012} or on Graphics Processing Units (GPU) \cite{Harada2007, sphcudafirst, Crespo2011} in order to accelerate SPH simulations. GPUs provide faster on-chip memory bandwidths compared to CPUs thanks to their architecture. Thus, GPU parallelization of the algorithms, where large blocks of data need to be processed is relatively more efficient compared to the CPU parallelization practices. Earlier studies using GPUs for SPH computations have utilized OpenGL Application Programming Interface (API) for parallelization \cite{Harada2007}. Using OpenGL for GPU parallelism requires conversion of mathematical operations to graphical rendering primitives, which may be quite cumbersome \cite{Du2012}. Following the introduction of CUDA API in 2007 by Nvidia Corporation, a more general-purpose interface for GPU computing is achieved. As a result, the utilization of CUDA is becoming increasingly popular in SPH research \cite{sphcudafirst, Crespo2011, Winkler2017, Chow2018, sphcuda} in parallel with the constant and rapid developments in the GPU architecture that aims to meet the needs of high-performance computing.

In this study, a novel implementation of the SPH method is presented for simulating the active nematics for the first time in literature to the best of authors' knowledge. To this end, a weakly compressible SPH algorithm is developed to be able to accurately discretize the governing equations of the active nematic. An in-house, object-oriented, and parallel computational tool is developed using the CUDA programming language based on the backbone of the extensively validated serial weakly compressible and incompressible SPH platform of our research group. Numerical simulations in an unconfined two-dimensional square domain are conducted where boundaries are treated as fully periodic through using ghost particles. The length scale of active nematic is investigated in detail through vorticity-vorticity correlation while the velocity scale of the system is explored by means of root mean square (rms) of velocity. The SPH results are extensively validated and verified with the results of relevant studies in literature in terms of vorticity and streamline structures as well as  vorticity-vorticity correlation. It is shown that present results are in a convincingly well agreement with those of literature and also with the other numerical methods. To shed light on the energy cascading of the nemeatic fluid, kinetic energy spectrum is analyzed with respect to the wavenumber at very low Reynolds number. It is also shown that due to its Lagrangian nature, the SPH method can readily and naturally capture formation of all topological defects as well as active walls on which these defects are located. The effect of extensile and contractile particles with the different activities as well as the elastic constant on the flow charateristics is scrutinized. Moreover, to further elucidate the advantages of the SPH method associated with its Lagrangian nature over the other numerical techniques used for modeling active nematics, the flow domain is divided into four quarters through color function, hence investigating the mixing behavior of nemeatic regions and the evolution of their initial interfaces. Finally, the entire history of the motion of five nematic particles is tracked so that it becomes possible to investigate the influence of the activity and elastic constant on the particle pathlines and the distance traveled by them. It is shown that the activity intensifies the chaotic nature of the active nematic by increasing the pathline and mixing efficiency whereas the elastic constant acts in a reverse manner with respect to the activity. This study which suggests a new perspective to the modeling of active nematics is organized as follows: \secref{sec:gov} introduces the continuum-based nematohydrodynamics model and governing equations of active nematics. The SPH method is introduced in \secref{sec:sph} along with the SPH discretization of governing equations. \secref{sec:cuda} discusses the parallel implementation of the proposed SPH model for nematohydrodynamics based on a CUDA platform and the results are presented in \secref{sec:res}. Finally, concluding remarks are provided in \secref{sec:conc}.

\section{\label{sec:gov}Governing equations}
In this study, we have developed an active nematohydrodynamics model \cite{nematiccorrelation1,nematiccorrelation2} using the SPH method to investigate the hydrodynamics of incompressible active nematics through utilizing the Beris-Edwards model (BE). As discussed in \secref{sec:int}, BE model has been mathematically well studied approach based on the Landau-de Gennes theory and a set of partial differential equations and has been shown to be successful in describing active nematics based on the continuum perspective \cite{amin2018}. To describe the thermodynamics of phase transition in liquid crystals, one need to introduce the director field and at least, one additional structural parameter to evaluate the degree of alignments. Early attempts utilized the $cos(\theta)$ function for this evaluation. Nevertheless, this approach was not successful for the nematics due to the fact that nematic particles have head-tail symmetry structure which means $\veb{n}$ and $-\veb{n}$ are equivalent thereby leading to the same ordered state (see \figref{fig:schematics}) \cite{beristhermodynamics}. It should be noted that, as mentioned earlier, $\veb{n}$ is an apolar quantity (i.e., $\veb{n}$ = $-\veb{n}$ in vectorial notation). Furthermore, in the absence of any external forces and wall effects, the direction of $\veb{n}$ is arbitrary in the space and it contains the information about the local alignment of the nematic particles only, while it lacks the information about how well these alignments are. To rectify these deficiencies, a traceless and symmetric tensorial ordering parameter $\veb{Q}$ is introduced which is defined as $\veb{Q}=\frac{d}{d-1}q(\veb{n}\veb{n}-\veb{I}/d)$ where $d$ and $\veb{I}$ are the dimension of space and identity matrix, respectively. Parameter $q$ is the magnitude of the order which contains the information about the quality of particles alignment such that $q=1$ and $q=0$ respectively correspond to perfect orientational order and complete disorder. Essentially, in the BE model, the conservation of linear momentum for incompressible active nematics with anisotropic forces is mutually coupled with an advection-diffusion equation for the  $\veb{Q}$ tensor, which is also referred to as nematodynamic equation. The coupled solution of conservation of mass and linear momentum and Nematodynamic  equation provide respectively, velocity and pressure fields of the incopressible flow, and the evolution of the liquid crystal director field in the computational domain. The coupled solution enables the computation of complex interactions between the fluid and liquid crystals such that the flow fields affect the position and director field of the liquid crystals and correspondingly, the change in these attributes of the liquid crystal influence the flow fields. In passing, it should be noted that hereafter, vector and tensor quantities are respectively represented with  upper-case and lower-case bold letters.

The advection part of the nematodynamic equation for $\veb{Q}$ can be derived by taking the material time derivative of the director field $\veb{n}$, which can be shown to be of the form, $\DU\veb{n}/\DU t = \veb{n} \cdot \veb{\Omega} = \veb{n} \cdot(\grad \veb{u} - \veb{E}),$ where $\veb{\Omega}=\frac{1}{2}\crsc{\grad \veb{u} - (\grad \veb{u})^\dag}$ and $\veb{E}=\frac{1}{2}\crsc{\grad \veb{u} + (\grad \veb{u})^\dag}$ are the vorticity and the rate of strain tensors, respectively. The dag superscript $(\dag)$, $(\cdot)$ and $\grad$ respectively represent the transpose, inner dot product and the Nabla operators. Additionally, $\DU/\DU t$ represents the material time derivative which can be written as $\DU/\DU t=\partial/\partial t+\veb{u}\cdot\grad$. After performing some tedious mathematical manipulations, the time evolution of $\veb{Q}$ can be written as \cite{beristhermodynamics}:
\begin{equation}
	\frac{\DU \veb{Q}}{\DU t} -\veb{S} = \Gamma\veb{H},
	\label{eq:nematodynamic}
\end{equation}
where $\veb{S}$ accounts for generalized nonlinear convective term and defined as:
\begin{equation}
	\veb{S}=(\lambda \veb{E}+\veb{\Omega})\cdot(\veb{Q}+\veb{I}/3)+(\veb{Q}+\veb{I}/3)\cdot(\lambda \veb{E}-\veb{\Omega})-2\lambda(\veb{Q}+\veb{I}/3)(\veb{Q} : \grad \veb{u} ),
	\label{eq:corotation}
\end{equation}

in which, the operator $(:)$ is the double inner product between two tensorial fields, the parameter $\Gamma$ is the rotational diffusivity which accounts for the macroscopic elastic relaxation time of the orientation field $\veb{Q}$, and $\lambda$ is tumbling parameter that adjusts the alignment of nematics with the flow. Given that the two term together on the left hand side of Eq. \eqref{eq:nematodynamic} corresponds to objective time rate of $\veb{Q}$ tensor, by attributing different numerical values to the parameter $\lambda$, one may obtain disparate forms of objective time rates of the $\veb{Q}$ tensor field such that the upper and lower convected derivatives can be achieved by substituting $\lambda = 1$ and $\lambda = -1$, respectively. Moreover, $\lambda = 0$ corresponds to the corotational time derivative \cite{nematiccorrelation2}. Physically, $|\lambda| > 1$ and $|\lambda| < 1$ respectively correlate with flow alignment and flow tumbling of nematics under the shear effect \cite{Carenza2019}. The last term in the Eq. \eqref{eq:nematodynamic}, i.e., $\veb{H}$ is the molecular field tensor which describes the relaxation of $\veb{Q}$ tensor field and is computed through variational derivative of the free energy as

\begin{equation}
	\veb{H}=-\frac{\delta \mathcal{F}}{\delta \veb{Q}}+\frac{\veb{I}}{3}\mathrm{Tr}(\frac{\delta \mathcal{F}}{\delta \veb{Q}}),
	\label{eq:molecularfield}         
\end{equation}

in which $\mathrm{Tr()}$ stands for the trace operator, and $\mathcal{F}=\mathcal{F}_e+\mathcal{F}_b$, is the free energy. It should be noted that the deformation in the orientation is not free and it happens at the expense of energy. This free energy is obtained by adding two distinct energy sources, i.e., $\mathcal{F}_e$ and $\mathcal{F}_b$, corresponding to the elastic and bulk (or Landau-de Gennes) free energies, respectively. The former can be defined as \cite{demus2011handbook}:

\begin{equation}
	\mathcal{F}_e=\frac{k_1}{2}(\grad\cdot\veb{n})^2+\frac{k_2}{2}\crsc{\veb{n}\cdot(\grad\times\veb{n})}^2+\frac{k_3}{2}\crsc{\veb{n}\times(\grad\times\veb{n})}^2,
	\label{eq:elasticergy}
\end{equation}
where $k_1$, $k_2$ and $k_3$, respectively, accounts for the splay, twist and bend of nematic particles such that $k_1 \approx k_3 > k_2$ \cite{demus2011handbook} (elastic constant for the splay and bend have comparable magnitude and the twist is usually smaller than those). It is common to simplify \eqref{eq:elasticergy} by assuming the elastically isotropic medium where all elastic constants are equal $k_1=k_2=k_3=K$ \cite{dephysics}. Finally, the resultant equation is mapped such that $\veb{n}$ is replaced by $\veb{Q}$, i.e. $\mathcal{F}_e=\frac{K}{2}(\grad\veb{Q})^2$ \cite{dephysics}. The bulk free energy $\mathcal{F}_b$ is the chemical potential function which describes the equilibrium state of the nematics. At the high temperature, the minimum of $\mathcal{F}_b$ corresponds to the isotropic state ($\veb{Q}=0$). For the low temperatures, the minimum energy occurs at the point in which any two eigenvalues of $\veb{Q}$ are equal \cite{mottram2014introduction}. According to the Landau-de Gennes theory, for the nematic particles, this function can be written by using the Taylor expansion for $\veb{Q}$ as \cite{dephysics}:

\begin{equation}
	\mathcal{F}_b=\frac{A}{2}\veb{Q}^2+\frac{B}{3}\veb{Q}^3+\frac{C}{4}\veb{Q}^4+\mathcal{O}(\veb{Q}^5),
	\label{eq:bulkenergy}
\end{equation}
where, $A$, $B$, and $C$ are the material parameters \cite{dephysics}. Finally the free energy reads:
\begin{equation}
	\mathcal{F}=\frac{A}{2}\veb{Q}^2+\frac{B}{3}\veb{Q}^3+\frac{C}{4}\veb{Q}^4+\frac{K}{2}(\grad\veb{Q})^2.
	\label{eq:freeenergy}
\end{equation}
Combination of Eqs. \eqref{eq:freeenergy} and \eqref{eq:molecularfield} yields:
\begin{equation}
	\veb{H}=-A\veb{Q}+B\big(\veb{Q}^2-(\veb{I}/3)\mathrm{Tr}\veb{Q}^2\big)-\veb{Q}\mathrm{Tr}\veb{Q}^2+K\grad^2\veb{Q}.
	\label{eq:molecularfielddis}
\end{equation}

Assuming a Newtonian fluid, velocity field is governed by the conservation of mass and linear momentum equations as follows:

\begin{equation}
	\frac{\DU \rho}{\DU t} = -\rho \grad \cdot \veb{u},
	\label{eq:mass}
\end{equation}
\begin{equation}
	\rho \frac{\DU \veb{u}}{\DU t} = \grad \cdot \boldsymbol{\Pi},
	\label{eq:mom}
\end{equation}
where $\veb{u}$, $\rho$, and $t$ are the velocity vector, density, and time respectively. It should be noted that although Eq. \eqref{eq:mass} can be simplified to $\grad \cdot \veb{u} = 0$ for the incompressible fluid, we will abide by the general compressible form since the incompressible flow is approximated by a weakly compressible scheme in the SPH method. Furthermore, $\boldsymbol{\Pi}$ denotes the general stress tensor, 
\begin{equation}
	\boldsymbol{\Pi}=\Pi_{viscous}+\Pi_{elastic}+\Pi_{active},
	\label{eq:stress}
\end{equation}
where $\boldsymbol{\Pi}_{viscous}$, $\boldsymbol{\Pi}_{elastic}$, and $\boldsymbol{\Pi}_{active}$ are viscous, elastic, and active stresses, respectively: 
\begin{equation}
	\boldsymbol{\Pi}_{viscous}=2 \mu \veb{E},
	\label{eq:viscousstress}
\end{equation}

\begin{equation}
	\boldsymbol{\Pi}_{elastic}=-p\veb{I}+2\lambda(\veb{Q}+\veb{I}/3)(\veb{Q}:\veb{H}-\lambda\veb{H}\cdot(\veb{Q}+\veb{I}/3)-\lambda(\veb{Q}+\veb{I}/3)\cdot\veb{H},
	\label{eq:elasticstress}
\end{equation}

\begin{equation}
	\boldsymbol{\Pi}_{active}=-\zeta \veb{Q}.
	\label{eq:activestress}
\end{equation}
In the above equations, $\mu$ and $p$ denote viscosity and pressure, respectively, and $\zeta$ is activity parameter, where $\zeta>0$ and $\zeta<0$ correspond,
respectively, to extensile (pusher) and contractile (puller) particles.

It should be noted that as mentioned in \cite{Thampi_2015,amin2018},
for systems of microscopic particles, as in active nematics, due to their small size and velocities, the governing equations can be significantly simplified. For such systems, $\boldsymbol{\Pi}_{elastic}$ is dominated by active stress \cite{Giomi2015}. Consequently, some of terms in RHS of \eqref{eq:stress} can be neglected. Thus, it becomes as follows:

\begin{equation}
	\boldsymbol{\Pi}=-p\veb{I}+\boldsymbol{\Pi}_{viscous}+\boldsymbol{\Pi}_{active}.
	\label{eq:mainstress}
\end{equation}

This equation along with equations \eqref{eq:nematodynamic}, \eqref{eq:mass}, and \eqref{eq:mom} specify the hydrodynamics of active nematic fluid in our simulation.

\section{\label{sec:sph}Numerical Method}

An exact integral representation of a function $f$ on the set of coordinates $x,x_0\in\mathbb{R}$ can be obtained by the following expression;

\begin{equation}\label{int_exact}
	f(x_0)=\int_{-\infty}^{\infty}f(x)\delta(x_0-x)dx,
\end{equation}
where $\delta(x_0-x)$ is the Dirac delta function that has the properties
\begin{equation}\label{Dirac}
	\delta(x_0-x)=
	\begin{cases}
		0,		& x\neq x_0\\
		\infty,	& x=x_0
	\end{cases}
	,\qquad \int_{-\infty}^{\infty}\delta(x_0-x)dx=1,\qquad \delta(x_0-x)=\delta(x-x_0).
\end{equation}
%We can delete the part above.
In SPH, spatial discretization of the continuum is actualized via moving particles, which can be considered as Lagrangian interpolation points. These particles, as they move in space and time, are able to carry the properties of the material that they represent. Therefore, on the constantly evolving ensemble of particles that compose the domain, value of any arbitrary field function can be approximated by the SPH kernel approach as

\begin{equation}\label{int_sph}
	f(\mathbf{r_i})\equiv\int_{\Omega}f(\mathbf{r_j})W(\mathbf{r_i}-\mathbf{r_j},h)d^3\mathbf{r_j},
\end{equation}
where $f(\mathbf{r_i})$ represents the value of the function $f$ on the spatial coordinate set denoted by the vector $\mathbf{r_i}\in\mathbb{R}^3$ for the particle of interest $\mathbf{i}$. Here, $d^3\mathbf{r_j}$ is the differential particle volume in the continuum. Bounded volume of the integral on the RHS of Equation \eqref{int_sph} is defined by the compact support domain $\Omega$, whereas $\mathbf{r_j}$ denotes neighboring coordinates of particle $\mathbf{i}$. The foundation of the SPH method is based on the kernel function $W(\mathbf{r_i}-\mathbf{r_j},h)=W(\mathbf{r_{ij}},h)=W_\mathbf{ij}$, which is an approximated form of $\delta(\mathbf{r_i}-\mathbf{r_j})$. Here in short notation, $\mathbf{r_{ij}}$ is equal to $\mathbf{r_i}-\mathbf{r_j}$. Equation \eqref{int_sph} dictates an interpolation within a spherical neighborhood defined by the smoothing length parameter $h$, in which the kernel function acts as a weighting factor. 
In this study, the quintic kernel function is utilized, which reads
\begin{equation}\label{quintic}
	W(b,\kappa h)=a_d
	\begin{cases}
		(3-b)^5-6(2-b)^5+15(1-b)^5,		& 0\leq b\leq 1 \\
		(3-b)^5-6(2-b)^5,				& 1\leq b\leq 2 \\
		(3-b)^5,						& 2\leq b\leq 3 \\
		0,								& b\geq 3 
	\end{cases}
	.
\end{equation}
Here, $a_d$ is the kernel normalization factor which is defined as $120/h$, $7/(478\pi h^2)$ and $3/(359\pi h^3)$, respectively, for one, two and three dimensions. The argument $b$ is defined as $b=r_\mathbf{ij}/h$ and $r_\mathbf{ij}=\left\|\mathbf{r_i}-\mathbf{r_j}\right\|$, while $\kappa$ is the coefficient that extends the smoothing length in accordance with the conditional form of the kernel function, which is taken as 3. In this study, $h$ is taken as $4/3$ times the initial particle distance in all simulations, which leads to an average of $\approx 45$ neighbors for each particle within a radius of $3h$.

So far, we used direct notation by which the vectors are represented with lower case bold-faced letters, whereas the tensors are denoted by upper case bold-faced letters. For convenience, we additionally introduce the index notation and utilize either index or direct notation for the remainder of the paper. When index notation is adopted, vector or tensor components will be denoted by italic Latin indices as superscripts, where repeating indices indicate summations over them as per the Einstein summation convention.

In a discretized particle set-up, where $N$ represents the number of neighbors in the support domain of particle $\mathbf{i}$, Equation \eqref{int_sph} can be expressed as
\begin{equation}\label{sphSum}
	f_{\mathbf{i}}^{s}=\sum_{\mathbf{j}=1}^{N}V_{\mathbf{j}}f_{\mathbf{j}}^{s} W_{\mathbf{ij}}.
\end{equation}
Here, the value of function $f_{\mathbf{i}}^{s}$ is either a scalar, vector or tensor, and the variable $V_\mathbf{j}$ denotes the volume of particle, which replaces the term $d^3\mathbf{r_j}$ in Eq.\eqref{int_sph} and calculated for each particle as $V_{\mathbf{i}}=1/\sum_{\mathbf{j}=1}^{N} W_\mathbf{ij}$.

On the other hand, first and second-order spatial derivatives of a function can be computed by applying the SPH particle approach \eqref{sphSum} on the Taylor series expansion and also benefiting from the properties of a second-rank isotropic tensor, leading to the corrective SPH formulation, which is explained in detail in \cite{Shadloo2011} and reads as follows

\begin{equation}
	\frac{\partial f_{\mathbf{i}}^{s}}{\partial x_{\mathbf{i}}^{k}}\alpha_{\mathbf{i}}^{kl}=\sum_{\mathbf{j}=1}^{N}V_{\mathbf{j}}\left ( f_{\mathbf{j}}^{s}-f_{\mathbf{i}}^{s} \right )\frac{\partial W_{\mathbf{ij}}}{\partial x_{\mathbf{i}}^{l}},
	\label{eq:sph1stDerivative}
\end{equation}

\begin{equation}
	\frac{\partial}{\partial x_{\mathbf{i}}^{k}}\left (\frac{\partial f_{\mathbf{i}}^{s}}{\partial x_{\mathbf{i}}^{k}}  \right )\alpha_{\mathbf{i}}^{sl}=8\sum_{\mathbf{j}=1}^{N}V_{\mathbf{j}}\left ( f_{\mathbf{i}}^{s}-f_{\mathbf{j}}^{s} \right )\frac{r_{\mathbf{ij}}^{s}}{r_{\mathbf{ij}}^{2}}\frac{\partial W_{\mathbf{ij}}}{\partial x_{\mathbf{i}}^{l}}.
	\label{eq:sphVectorLaplacian}         
\end{equation}	
Eq. \eqref{eq:sph1stDerivative} and Eq. \eqref{eq:sphVectorLaplacian} are first and second-order derivatives of a function, respectively. The corrective SPH approach is proved to be vastly beneficial in reducing numerical errors originated from irregular particle distribution and also from truncated support domains in boundary regions. Here, $\alpha_{\mathbf{i}}^{sl}$ is the second rank correction tensor, which is basically the uncorrected first-order spatial derivative of particle position:

\begin{equation}\label{sphCorrectionMatrix}
	\alpha_{\mathbf{i}}^{sl}=\sum_{\mathbf{j}=1}^{N}r_{\mathbf{ji}}^{s}V_{\mathbf{j}}\frac{\partial W_{\mathbf{ij}}}{\partial x_{\mathbf{i}}^{l}}.
\end{equation}	

In the classical weakly compressible SPH (WCSPH) scheme \cite{Gingold_Monaghan}, fluid flow is assumed to be weakly compressible by allowing particle densities to vary within a limit of $1\%$ \cite{Monaghan1994}. Using this limited compressiblity approach, pressure field is linked with the density variations of the particles, which yields an explicit time integration algorithm for the solution of the governing equations of the fluid flow. Therefore, Eq. \eqref{eq:mom} can be discretized using above corrective SPH approximations as

\begin{equation}\label{eq:sphMomentum}
	\frac{D\mathbf{u}_{\mathbf{i}}}{Dt}=-\rho_{\mathbf{i}}\sum_{\mathbf{j}=1}^{N}V_\mathbf{j}\left ( \frac{p_{\mathbf{i}}}{\rho_\mathbf{i}^2}+\frac{p_{\mathbf{j}}}{\rho_\mathbf{j}^2} \right ) (\mathbf{B}_\mathbf{i}\cdot \grad_\mathbf{i}W_{\mathbf{ij}})+8\nu\frac{\rho_0}{\rho_\mathbf{i}}\sum_{\mathbf{j}=1}^{N}V_\mathbf{j}\frac{\left ( \mathbf{u}_\mathbf{i}-\mathbf{u}_\mathbf{j} \right )\cdot\left ( \mathbf{r}_\mathbf{i}-\mathbf{r}_\mathbf{j} \right )}{\left \|\mathbf{r}_\mathbf{i}-\mathbf{r}_\mathbf{j}  \right \|^2}(\mathbf{B}_\mathbf{i}\cdot \grad_\mathbf{i}W_{\mathbf{ij}})+\sum_{\mathbf{j}=1}^{N}V_{\mathbf{j}}\left ( \veb{Q}_{\mathbf{j}}-\veb{Q}_{\mathbf{i}} \right )(\mathbf{B}_\mathbf{i}\cdot \grad_\mathbf{i}W_{\mathbf{ij}}),
\end{equation}
where $\rho_\mathbf{i}$ is the density of the particle, $\rho_0$ is reference fluid density, $\nu$ is the kinematic viscosity, and kernel normalization tensor $\mathbf{B_i}$ is the inverse of $\alpha_{\mathbf{i}}^{sl}$ in the direct notation form \cite{Kolukisa2020}. The first term on the right-hand-side of the Eq. \eqref{eq:sphMomentum} is the pressure force term, which is discretized with a symmetric SPH approach \cite{Liu2003} rather than employing Eq. \eqref{eq:sph1stDerivative} to compute pressure gradient. The time rate of change of ordering parameter, $Q$ is calculated via Eq. \eqref{eq:nematodynamic} for all particles as: 

\begin{equation}
	\begin{split}
		\frac{D\mathbf{Q}_{\mathbf{i}}}{Dt}=(\lambda \veb{E}_\mathbf{i}+\veb{\Omega}_\mathbf{i})\cdot(\veb{Q}_\mathbf{i}+\veb{I}/3)+(\veb{Q}_\mathbf{i}+\veb{I}/3)\cdot(\lambda \veb{E}_\mathbf{i}-\veb{\Omega}_\mathbf{i})-2\lambda(\veb{Q}_\mathbf{i}+\veb{I}/3)(\veb{Q}_\mathbf{i} : \grad \veb{u}_\mathbf{i}) \\
		+\Gamma\left[-A\veb{Q}_\mathbf{i}+B\big(\veb{Q}_\mathbf{i}^2-(\veb{I}/3)\mathrm{Tr}\veb{Q}_\mathbf{i}^2\big)-C\veb{Q}_\mathbf{i}\mathrm{Tr}\veb{Q}_\mathbf{i}^2+\grad^2\veb{Q}_\mathbf{i}\right].
		\label{eq:mi}
	\end{split}
\end{equation}
It should be noted that Eq. \eqref{eq:sph1stDerivative} and Eq. \eqref{eq:sphVectorLaplacian} are used, respectively, to calculate the gradient ($\grad$) and Laplacian ($\grad^2$) terms in Eq. \eqref{eq:mi}. Likewise, particle densities are updated according to the material derivative determined by the continuity equation (Eq. \eqref{eq:mass}) as:
\begin{equation}\label{sphContinuity}
	\frac{D\rho_\mathbf{i}}{Dt}=\rho_\mathbf{i}\sum_{\mathbf{j}=1}^N V_{\mathbf{j}}\left(\mathbf{u}_\mathbf{i}-\mathbf{u}_\mathbf{j}\right)\cdot\left( \mathbf{B}_\mathbf{i}\cdot \grad_\mathbf{i}W_{\mathbf{ij}} \right).
\end{equation}
In order to compute particle pressures in Eq. \eqref{eq:sphMomentum}, an Equation of State (EoS) is employed by the weakly compressible SPH method which reads:

\begin{equation}\label{eq:sphEos}
	p_\mathbf{i}=\frac{\rho_0c_0^2}{\gamma}\left[ \left( \frac{\rho_\mathbf{i}}{\rho_0} \right)^\gamma -1\right].
\end{equation}
Here, $c_0$ is the speed of sound parameter, and $\gamma$ is the specific heat-ratio, which is taken as 7. In this study, the value of $c_0$ is determined at each time step as 10 times of the maximum velocity in the domain in order to satisfy the incompressiblity condition, which is defined by the mach number as $M=(u/c)<0.1$. The EoS approach is the key to avoid the implicit solution for the momentum equation. However, it limits the time step size since the speed of sound parameter becomes the dominant velocity scale in the domain at the definition of the Courant-Friedrichs-Lewy (CFL) stability condition for WCSPH.

%\textcolor{red}{Sice the active nematic flow posses a small Reynolds number ($\rnl<<1$), velocity magnitude is very small. Hence, instead of solving Poisson's equation which is numerically expensive, it is preferred to solve Equation of State (EoS) to determine the particle pressures in Eq. \eqref{eq:sphMomentum}:}
%\textcolor{blue}{Since the active nematic flow possesses a small Reynolds number ($\rnl<<1$), it is more expedient to adopt the WCSPH approach, instead of utilizing numerically expensive fully incompressible schemes [cite] that require implicit solution for the pressure Poisson equation.}

%By the weakly compressible EoS approach, pressure field is linked with the density variatons of the particles. Therefore, momentum equation (Eq.\eqref{eq:mom}) can be solved explicitly.

%In this study, a modified Euler time integration scheme is implemented. 
\subsection{\label{sec:timeIntegration}Time integration and numerical treatments}
To complete the temporal transition, position, velocity, nematic order, and density of the particles are updated during the simulation by integrating the following relations at each time step:
\begin{equation}\label{sphTimeIntegrationDerivatives}
	\frac{D\mathbf{r}_\mathbf{i}}{Dt}=\mathbf{u}_\mathbf{i},\qquad \frac{D\mathbf{u}_\mathbf{i}}{Dt}=\mathbf{a}_\mathbf{i},\qquad
	\frac{D\mathbf{Q}_\mathbf{i}}{Dt}=\mathbf{m}_\mathbf{i},\qquad
	\frac{D\rho_\mathbf{i}}{Dt}=k_\mathbf{i}.
\end{equation}

The modified Euler predictor-corrector time integration scheme \cite{Ozbulut2014} begins with the projection of intermediate particle velocities and positions with half time step size as $\mathbf{u}_\mathbf{i}^{n+1/2}=\mathbf{u}_\mathbf{i}^n+0.5\mathbf{a}_\mathbf{i}^n\Delta t$ and $\mathbf{r}_\mathbf{i}^{n+1/2}=\mathbf{r}_\mathbf{i}^n+0.5\mathbf{u}_\mathbf{i}^{n+1/2}\Delta t$, respectively, where n is the temporal index and $\Delta t$ is the time step size. 

Due to the adopted time integration scheme, all the SPH interpolations to compute the material derivatives in Eqs. \eqref{eq:sphMomentum}, \eqref{eq:mi}, and \eqref{sphContinuity} are performed at this projected particle setup. Therefore, it is sufficient to perform a neighbor search and particle pairing operation only at this stage of the time integration procedure. After establishing the connections and computing the values of kernel function $W_{\mathbf{ij}}$, its gradient $\grad_\mathbf{i}W_{\mathbf{ij}}$, particle volumes $V_\mathbf{i}$, and all relevant pairing information between neighboring particles such as $\mathbf{r_{ij}}$ and $\mathbf{u_{ij}}$, the kernel normalization tensor $\mathbf{B_{i}}$ is also computed for each particle by the inverse of the correction tensor given by Eq. \eqref{sphCorrectionMatrix}.
Followingly, the continuity equation (Eq. \eqref{sphContinuity}) is solved for $k_\mathbf{i}^{n+1/2}$, and the particle densities are projected by half time step size as $\rho_\mathbf{i}^{n+1/2}=\rho_\mathbf{i}^n+0.5k_\mathbf{i}^{n+1/2}\Delta t$. Additionally, a density filtering treatment \cite{Ozbulut2020} is implemented in order to maintain a smooth spatial density distribution in the domain:

\begin{equation}\label{eq:densityCorrection}
	\widehat{\rho}_\mathbf{i}=\rho_\mathbf{i}-\beta\frac{\sum_{\mathbf{j}=1}^{N} \left( \rho_\mathbf{i}-\rho_\mathbf{j} \right)W_{\mathbf{ij}}}{\sum_{\mathbf{j}=1}^{N}W_{\mathbf{ij}}}
\end{equation}
where $\widehat{\rho}_\mathbf{i}$ is the corrected density and $\beta$ is a density smoothing coefficient which is used to eliminate spurious density variation, thereby enhancing the robustness of the algorithm without impairing the fidelity of the results. The value of $\beta$ varies between zero and unity, and the value of unity corresponds to well known Shepard's interpolation. In this study, $\beta$ is chosen to be equal to unity. Subsequently, the particle pressures are computed by the EoS (Eq. \eqref{eq:sphEos}). 
The evolution of the nematics field is ensured by solving the Eq. \eqref{eq:mi} for $\mathbf{m_i}^{n+1/2}$ and consequently updating the ordering parameter as $\mathbf{Q}_\mathbf{i}^{n+1}=\mathbf{Q}_\mathbf{i}^n+\mathbf{m}_\mathbf{i}^{n+1/2}\Delta t$.
%Here should be an explanation for the nematics ordering but we have to provide discretized (particle notation) formulations for Q.
Finally, the momentum equation (Eq. \eqref{eq:sphMomentum}) can be solved to obtain an estimated acceleration field at the midpoint of the time step as $\mathbf{a_i}^{n+1/2}$. Therefore, particle velocity, position, and densities can be corrected, respectively, as $\mathbf{u}_\mathbf{i}^{n+1}=\mathbf{u}_\mathbf{i}^n+0.5\mathbf{a}_\mathbf{i}^{n+1/2}\Delta t$, $\mathbf{r}_\mathbf{i}^{n+1}=\mathbf{r}_\mathbf{i}^{n+1/2}+0.5\mathbf{u}_\mathbf{i}^{n+1}\Delta t$, and $\rho_\mathbf{i}^{n+1}=\rho_\mathbf{i}^{n+1/2}+0.5k_\mathbf{i}^{n+1}\Delta t$. The Artificial Particle Displacement (APD) correction algorithm \cite{Shadloo2011} is applied upon the finalization of the time step, in order to eliminate instabilities that are originated from particle clustering and particle fractures:

\begin{equation}\label{APD}
	\delta\mathbf{r}_\mathbf{i}=\sum_{\mathbf{j}=1}^{N}\frac{\mathbf{r}_{\mathbf{ij}}}{r_\mathbf{ij}^3}r_0^2 u_{v}\Delta t
\end{equation}
Here, $\delta\mathbf{r}_\mathbf{i}$ is the corrected particle position, $u_v=\left|\sum_{\mathbf{j}=1}^{N}\left( \mathbf{u}_\mathbf{i}-\mathbf{u}_\mathbf{j} \right)W_{\mathbf{ij}}\right|/{\sum_{\mathbf{j}=1}^{N}W_{\mathbf{ij}}}$ is the velocity variance based APD coefficient \cite{Ozbulut2017} and $r_0=\sum_{\mathbf{j}=1}^{N}r_{\mathbf{ij}}/N$ is the average neighbor distance of the particle $\mathbf{i}$.

\begin{figure}[h!]
	\centering
	\includegraphics[width=0.75\linewidth]{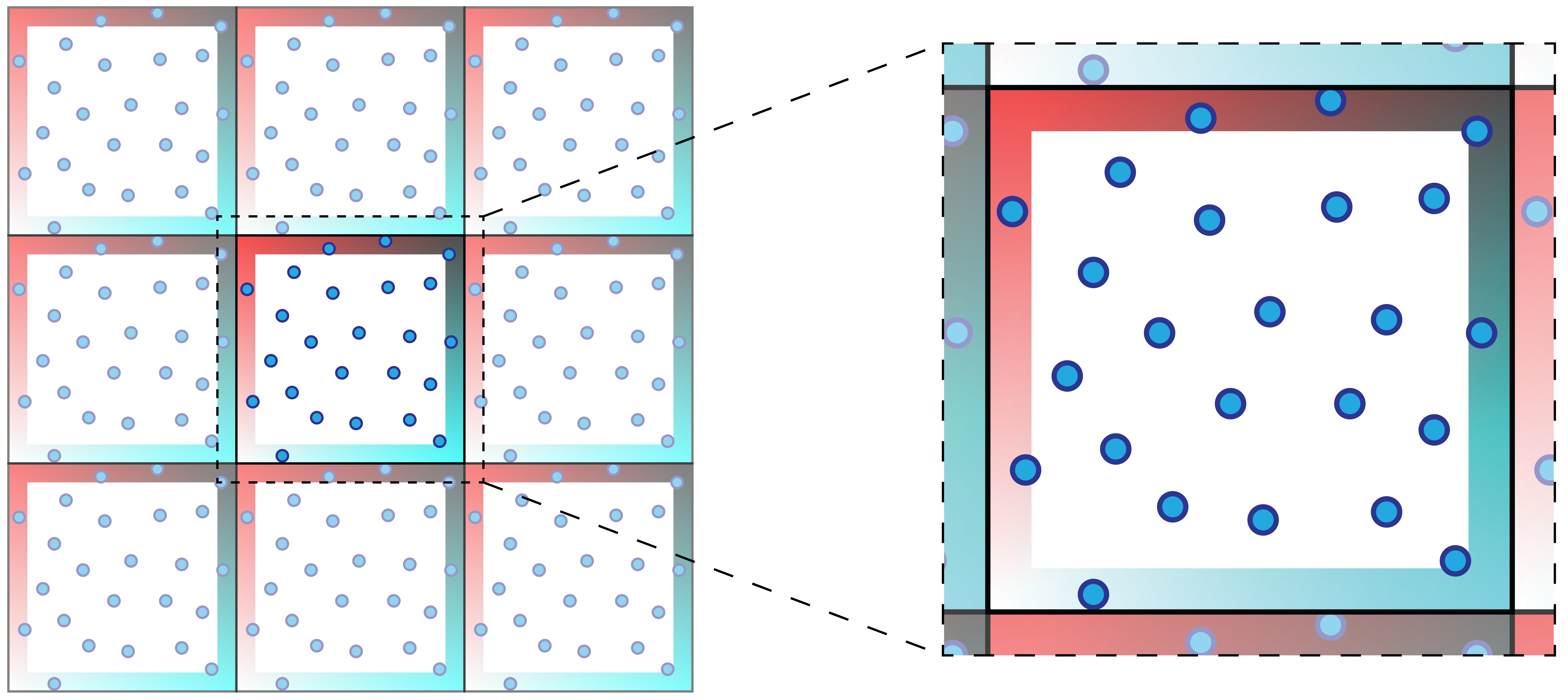}
	\caption{Distribution of ghost particles for periodic boundary condition. Color gradients depict unique spatial boundary regions for fluid particles in unit domain and their respective ghost copies.}
	\label{fig:boundary}
\end{figure}

%Boundary:
In this study, the geometry of the computational domain is a 2D square with a side length of $L$, where periodic boundary conditions apply in all directions to simulate an unconfined flow domain. The value of any arbitrary function $\phi(x,y,t)$ should satisfy the conditions;

\begin{equation}\label{boundaryconditions}
	\begin{split}
		\phi(0,y,t)=\phi(L,y,t),\qquad \phi(x,0,t)=\phi(x,L,t), \\
		\phi_x(0,y,t)=\phi_x(L,y,t),\quad \phi_y(x,0,t)=\phi_y(x,L,t).
	\end{split}
\end{equation}	
To implement this periodic boundary approach, the particles within the neighborhood ($\leq3h$) of boundary faces are copied and displaced, as depicted in Figure \ref{fig:boundary}. Therefore, any type of boundary effect is avoided without applying additional treatments on the boundary planes. This technique consequently prevents kernel truncation and assures continuity of the field functions on the boundaries. Furthermore, particles that leave the domain are displaced back into the domain with $\pm L$ in the direction of boundary normal for spatiotemporal consistency. Hence, conservation of mass is ensured in a fully periodic computational domain.

\section{\label{sec:cuda}GPU-based SPH algorithm}

\begin{figure}[hbt!]
	\centering
	\includegraphics[width=5.0in]{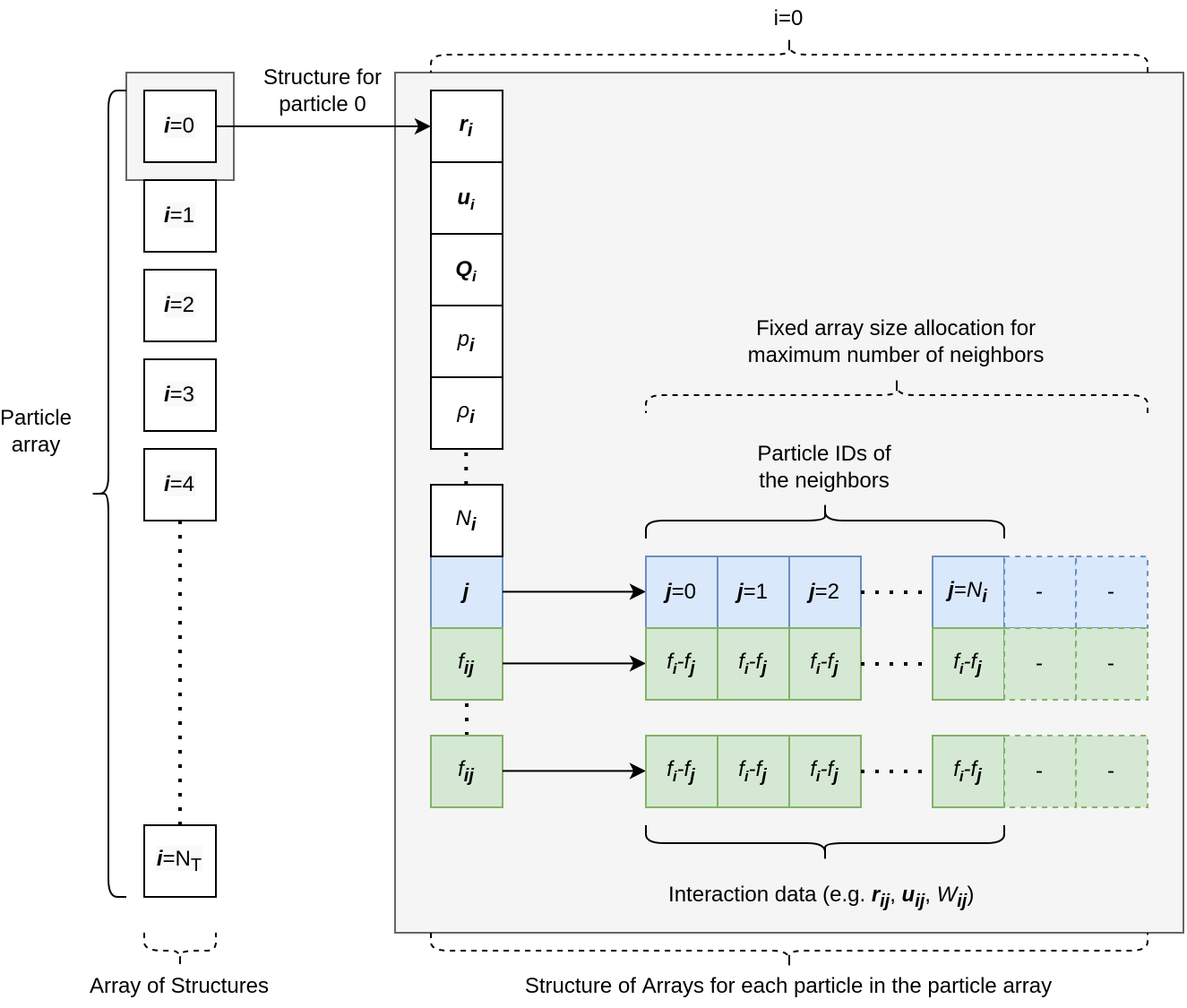}
	\caption{Array of Structures of Arrays organization for storing particle and interaction data. The column on the left hand side is the particle array. Each member of the particle array consists  of a SoA as depicted on the right hand side. The SoAs for each particle "$\mathbf{i}$" have members of variables that store function values for the given particle and have array members as well to store neighbor interaction data.}
	\label{fig:ParticleArray}
\end{figure}

An in-house computer code parallelized on the GPU, and based on CUDA and C++ programming languages is developed in order to perform the simulations of the study. An object oriented approach that focuses on flexibility and ease of programming rather than pure computational efficiency is adopted in the design of the data structures of the program, which have also enabled the integration of the nematics variables into the SPH particle framework effortlessly. To this end, particle data is organized based on the Array of Structures of Arrays (AoSoA) principle. Physical properties of particles, as well as their relative values between their neighbors are stored in a Structure of Arrays (SoA) for each particle, which are the elements of the particle array as depicted in Figure \ref{fig:ParticleArray}. Through this AoSoA approach, memory coalescence is obtained for the intermediary interaction variables (e.g. $\mathbf{r_{ij}}$, $\mathbf{u_{ij}}$, $W_{\mathbf{ij}}$) with repetitive usage.

%Due to the locality of SPH interactions, multiple component schemes require identification of neighboring particles.

Computations for each particle of the SPH system are assigned to a single thread within the framework of the Single Instruction Multiple Thread (SIMT) execution model of the CUDA programming language. These large numbers of threads are grouped as blocks of threads and are designed to work simultaneously in a parallel manner. All threads have access to the global memory of the GPU, while the threads within the same block also have access to a specific shared memory, and each thread has its own local memory. Following the initial distribution of particles, the particle array is copied from the host (CPU) memory to the device (GPU) global memory. Parallel computations are performed via the CUDA kernel functions, which are written in isolated serial forms, and executed simultaneously for each thread/particle in the SIMT framework. It should be noted that the term CUDA kernel is a programming concept that should be distinguished from the term kernel function in SPH formalism.

\begin{figure}[hbt!]
	\centering
	\includegraphics[width=5.0in]{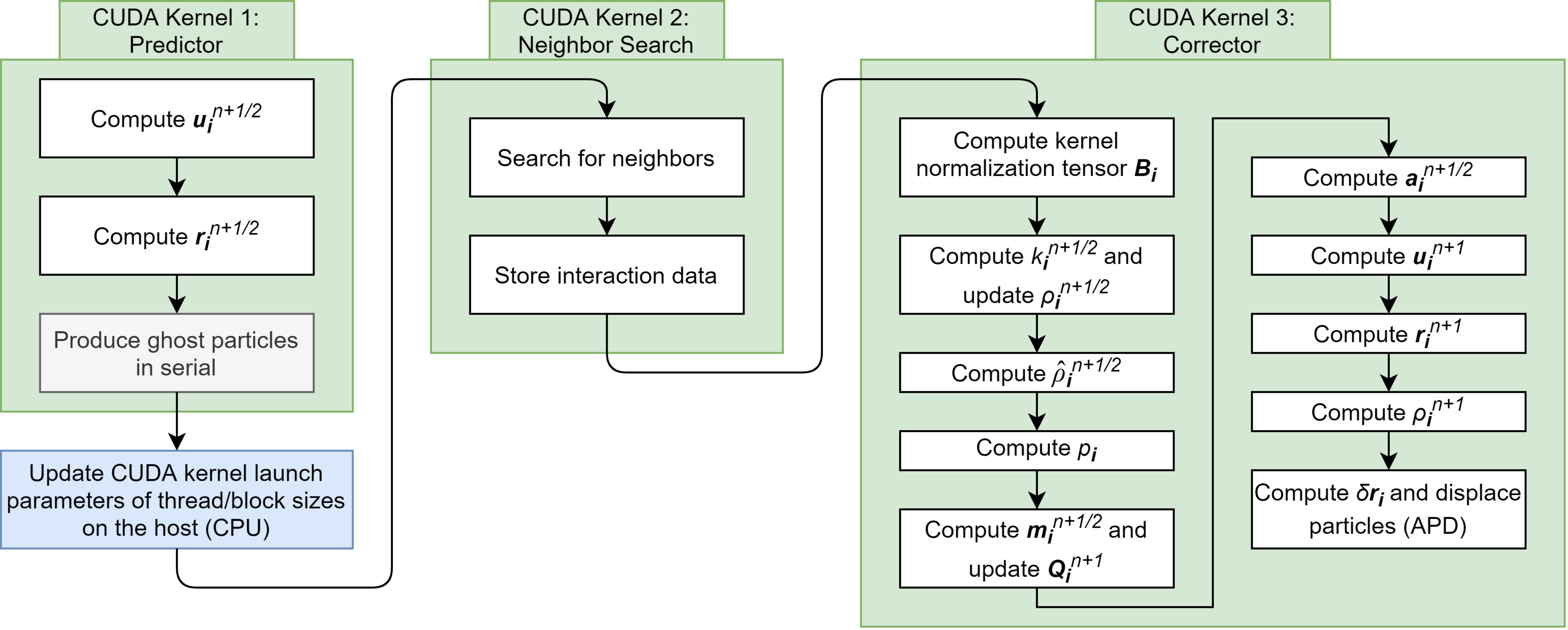}
	\caption{The algorithm for the computations of one time step}
	\label{fig:Algorithm}
\end{figure}

Generating ghost boundary particles are problematical for parallel algorithms, since a so-called race condition appears when boundary fluid particles attempt to insert their corresponding ghost duplicates into the particle array simultaneously. Therefore, ghost particle generation needs to be handled serially. In this study, a single GPU thread within a CUDA kernel is utilized for this task in order to avoid performing a costly two-way memory copying operation of particle array between the device and the host at each time step. An outline of the computational algorithm for one time step is schematized in Figure \ref{fig:Algorithm}. The sequential instructions which are grouped under CUDA kernels 1, 2, and 3 in Figure \ref{fig:Algorithm} are serial instructions for a single thread that is mapped for the computations of a single particle. However, the ghost particle production operation is an exception since it is performed serially by a single GPU thread. It should be noted that the computations that require summations of the neighbor values include nested serial loops within the thread. %%%%This last sentence is too obvious and may be deleted%%%%
A GPU-optimized neighbor searching algorithm \cite{Green2007} is implemented for computing neighbor interactions. In this approach, the flow domain is divided by square cells, and a pseudo particle number array is sorted according to their cell numbers at each time step. Subsequently, linked lists of the pseudo particle number arrays are created for each cell by utilizing shared memory arrays. As a result, a data structure is obtained, where the threads of each particle can easily search the linked lists of neighboring cells. In order to avoid performing this search operation repeatedly within a time step, particle IDs of neighboring particles are stored in an array for each particle. 

All main time-step computations are performed by the CUDA kernels. However, the particle array is only copied from device memory to the host memory when the program needs to write the outputs for the post-process. It should be noted that the neighbor interaction arrays, as well as the main particle array, have pre-allocated maximum sizes. Ghost particles are added after the end of the fluid particles on the array and their corresponding data are overwritten at each time step, while a variable holds the value of the actual particle number in order to define the end of the meaningful data. The same process applies to the particle neighbor interaction arrays, which also have variable sizes throughout the simulation.

\begin{table}
	\centering
	\caption{Performance profile of the CUDA kernels.}\label{tab:performance}
	\begin{tabular}{ c c } \toprule
		{Operation} & {Percentage of computational cost} \\ \midrule
		CUDA Kernel 1  & $43.90\%$ \\
		CUDA Kernel 2  & $30.71\%$ \\
		CUDA Kernel 3  & $24.08\%$ \\
		Other CUDA operations  & 1.31$\%$ \\ \bottomrule
	\end{tabular}
\end{table}

\begin{figure}[hbt!]
	\centering
	\includegraphics[width=5.0in]{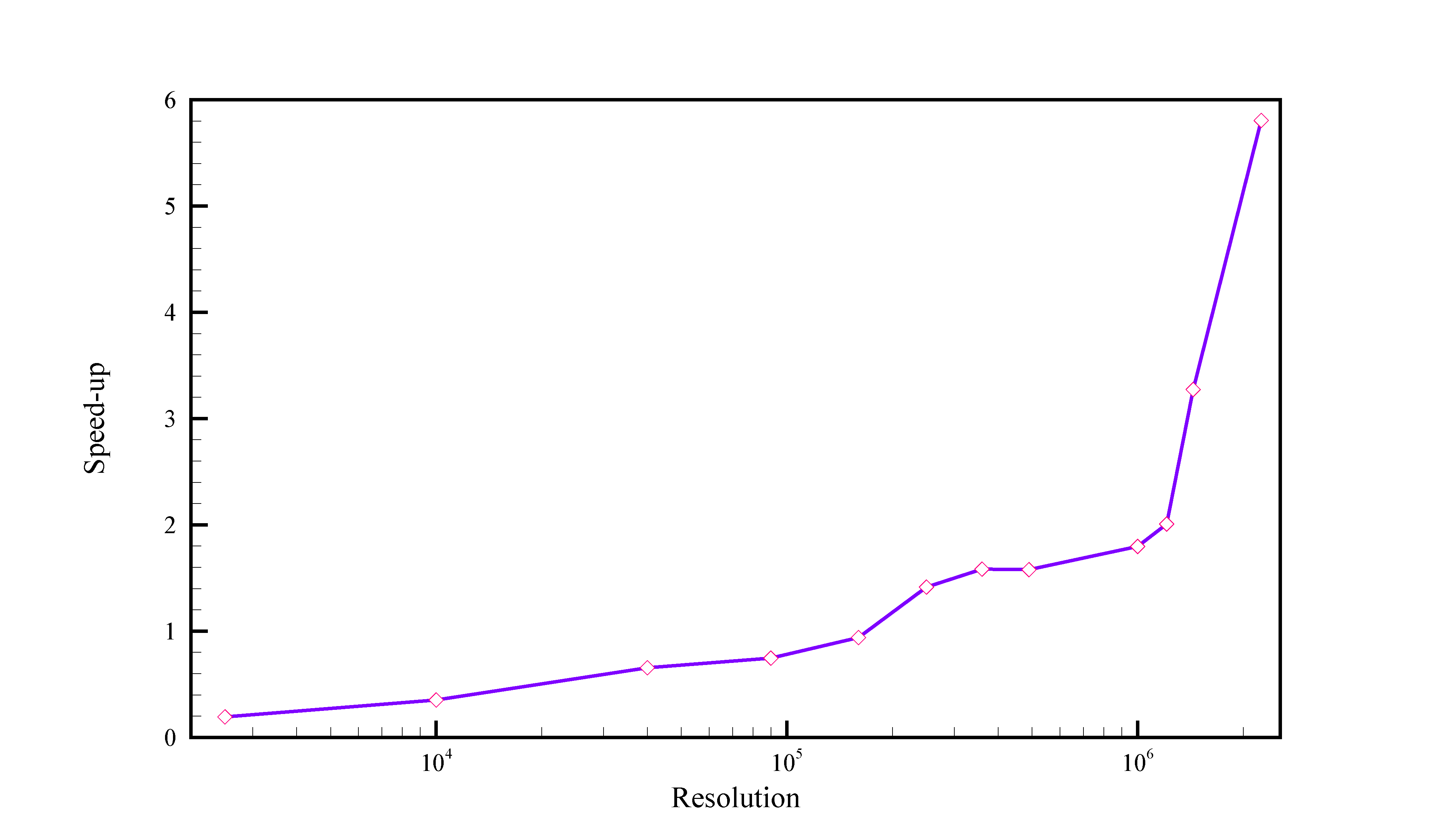}
	\caption{Speed-up of the current CUDA based SPH solver with respect to the serial OpenFOAM solver. Although the OpenFOAM requires smaller solution time at lower resolutions ($< O(5)$), significant speed-up is obtained by SPH for higher resolutions ($O(6)<$).}
	\label{fig:speedup}
\end{figure}

Percentages of average computational time costs of the different CUDA kernel groups (Fig. \ref{fig:Algorithm}) over 100 iterations with $\approx 2.5\times10^{5}$ particles are provided in Tab. \ref{tab:performance}. Percentage of the CUDA kernel 1 reveals the impact of the serial ghost particle algorithm on the performance of the program. Here, it can also be inferred that any improvement on the neighbor searching algorithm may also lead to a significant increase in the performance.
To demonstrate the speed-up achieved by the CUDA-based parallel algorithm of this study, the ratio of computational time required by the the serial Finite Volume solver of OpenFOAM \cite{OpenFOAM} to the time needed by the parallel SPH algorithm is plotted as a function of mesh resolution in Fig. \ref{fig:speedup} for the solution of the same problem. Generally, mesh-based algorithms are expected to be faster than mesh-free methods since they have fewer number of neighbors for each computational node and do not require updated neighbour lists throughout the simulation. As seen in this figure, the current GPU-parallelized SPH method notably outperforms the serial OpenFOAM particularly at higher resolutions although in this comparison, the time integration scheme used in the SPH method is second order Runge-Kutta whereas the one employed in the OpenFOAM is the first order Runge-Kutta (Euler) method.

\section{\label{sec:res}Numerical Simulation and results}

\begin{table}
	\centering
	\caption{The parameters used in the simulations.}\label{tab:dim}
	\begin{tabular}{ c c c } \toprule
		{Parameter} & {value} \\ \midrule
		$\Gamma$   & $0.4(Pa^{-1}s^{-1})$ \\
		$\lambda$  & $0.7$ \\
		$A$, $B$ and $C$  & $1$, $0$ and $0(Pa)$ \\
		$K$  & $0.02 (N)$ \\ 
		$\mu$  & $2/3 (Pas)$  \\ 
		$\rho$  & $1(kgm^{-3})$  \\ 
		$\zeta$ & $0.023(Pa)$ \\ \bottomrule
	\end{tabular}
\end{table}

The coupled governing equations \eqref{eq:nematodynamic}, \eqref{eq:mass}, and \eqref{eq:mom} are solved using the SPH method. The computational domain is a two-dimensional square one with a normalized size of unity. In order to avoid any confinement effect, a fully periodic unit domain is ensured by the periodic ghost boundary treatment presented in \secref{sec:timeIntegration}. The particle resolution is taken such that the total number of particles inside the domain is $\approx 2.5\times10^{5}$ which corresponds to the initial particle distance $\Delta x = 0.02 m$, and the time step size is determined according to the CFL condition.
With the aforementioned particle configuration, average computation time required for one time step iteration is $\approx0.5 s$ on an Nvidia Quadro RTX 5000 GPU.
The parameters used are provided in the Tab.\ref{tab:dim}, unless stated otherwise. Zero initial velocity is used with a slightly perturbed nematic orientation as the initial condition and the quantitative results are taken at the statistically steady state.

\begin{figure}[hbt!]
	\centering
	\includegraphics[width=4.5in]{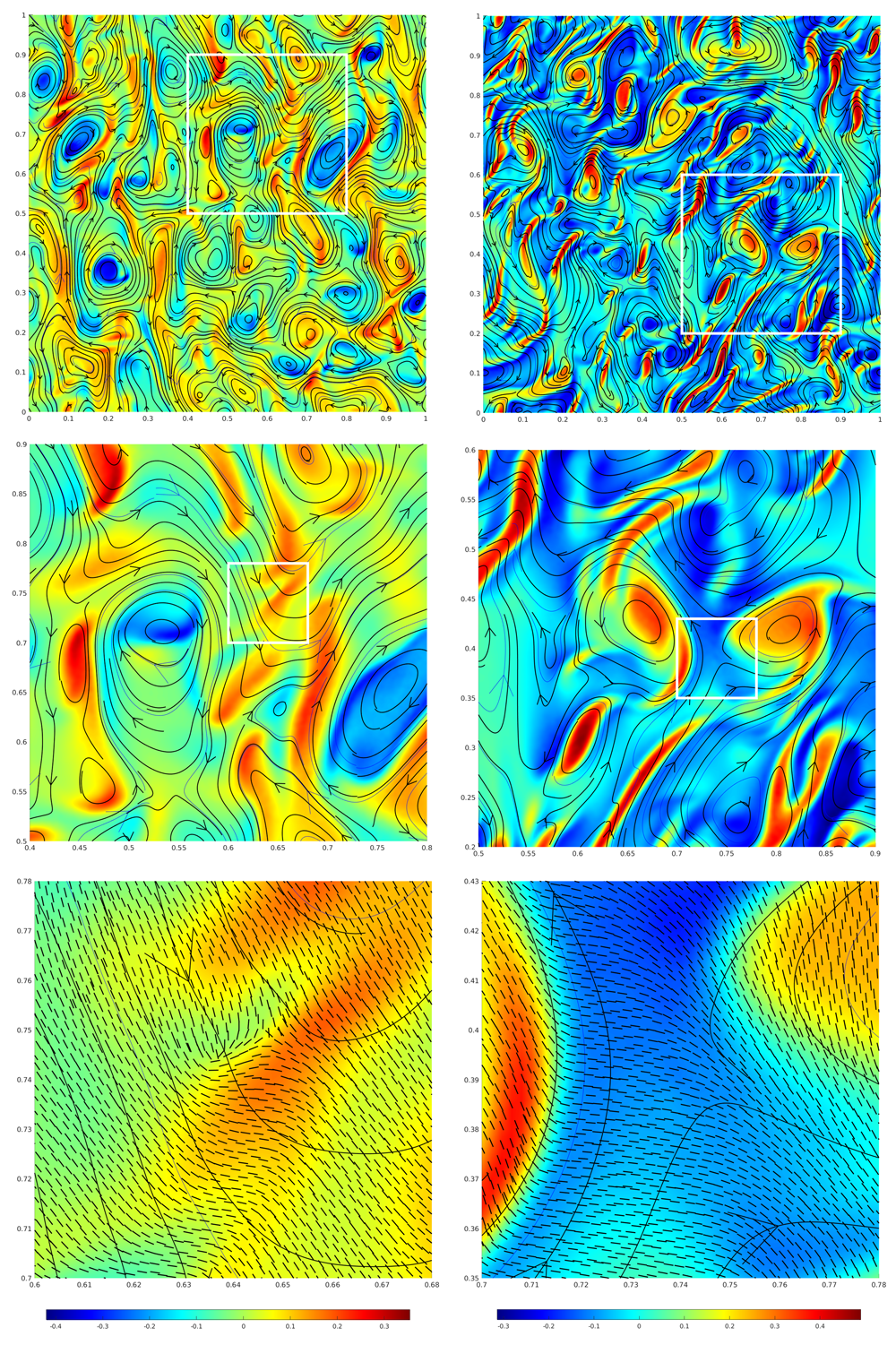}
	\caption{Vorticity contour and streamlines of the domain corresponding to $\zeta=0.03$ (left column) and $\zeta=-0.03$ (right column) at three different length scales. Lower subfigures corresponds to close up views and nematic directors are also shown only in last row.}
	\label{fig:vortex}
\end{figure}

As discussed in \secref{sec:int}, there is no source of external energy in active fluids, and energy is injected by the particles themselves. Since nematic fluids are inherently unstable \cite{Edwards_2009}, as the time progress, these instabilities reveal themselves in the form of vorticity.
In \figref{fig:vortex}, vorticity contours as well as streamlines are shown inside the computational domain for two different positive and negative activities at $\mu=0.1 pas.s$. These results are in good agreement with previous numerical and experimental studies (see for example figure 2 of \cite{nematiccorrelation1}). As can be seen in these figures, turbulent-like flow is created because of the existence of vortices inside the domain which consequently occurs due to the presence of activity that destroys the long-range nematic ordering. Particle alignments are calculated and shown in \figref{fig:vortex}-c. In this figure, the orientation of the nematic director is observed clearly and it is seen that the variation of vorticity depends on the nematic directors. 

\begin{figure}[hbt!]
	\centering
	\includegraphics[width=4.5in]{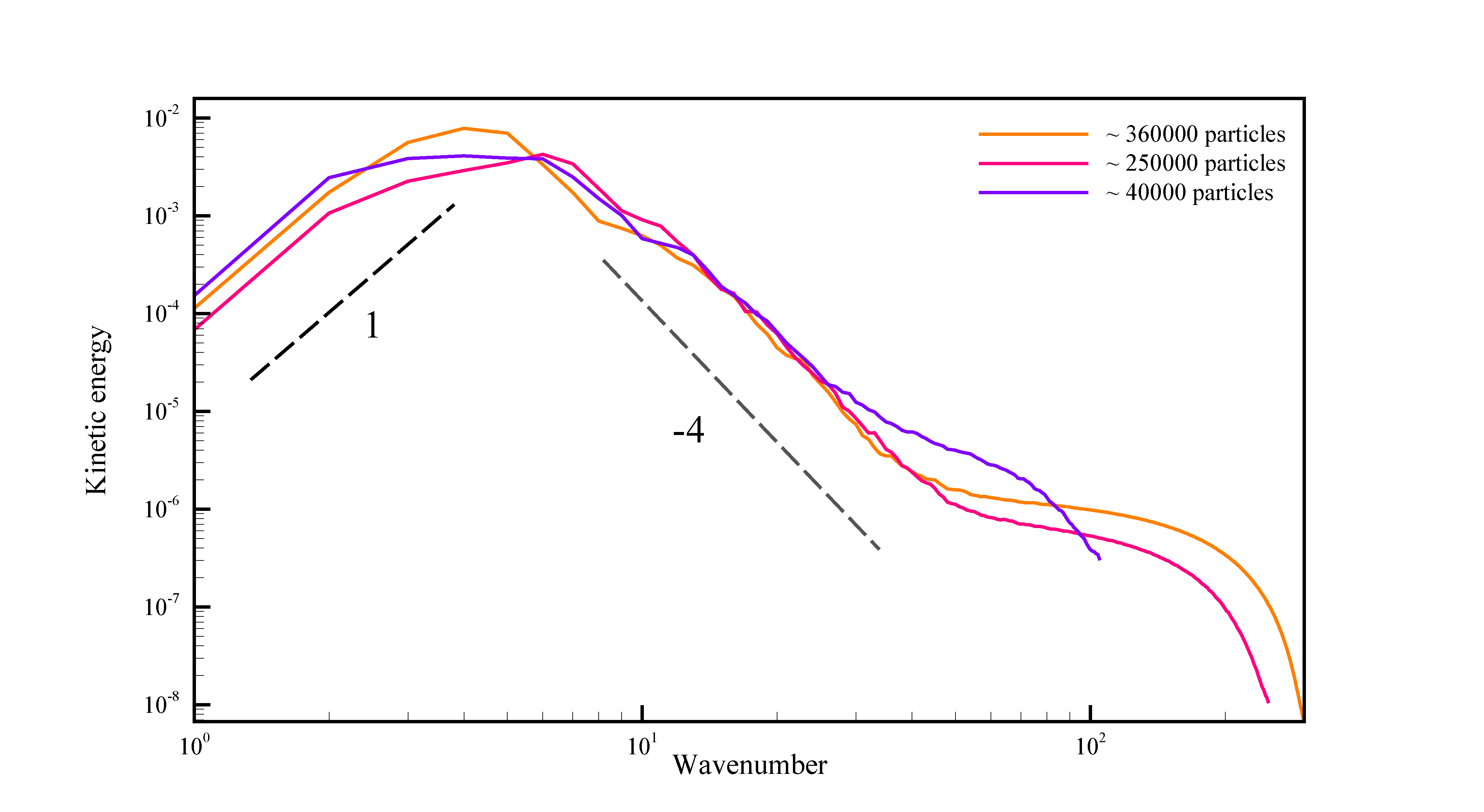}
	\caption{Spectrum of kinetic energy for different particle resolutions.}
	\label{fig:energy}
\end{figure}

It is common to use the kinetic energy per mass densityin Fourier mode, $E$, to analyze the structure of the turbulent flow:

\begin{equation}
	E=\frac{1}{2}\langle \mathbf{u}^2 \rangle,
	\label{eq:kineticenergy}
\end{equation}
where angular brackets $\langle . \rangle$ is the spatial average. In classic turbulent flow, a universal scaling was suggested by Kolmogorov \cite{Kolmogorov} as $E(w)\approx w^{-5/3}$, where $w=2\pi/l$ is the wavenumber. In the active fluid, however, in the absence of an external source, energy is injected by the active term.  Many researchers have tried to study the kinetic energy in the active fluid. Most recently, Alert \etal \cite{universalturbulence} proposed a universal scaling for the active nematics, yet disregarding the topological defects. In the present study, kinetic energy is calculated for different resolutions, and the results are presented in \figref{fig:energy}. In the active fluids, the energy is injected at the wide ranges of length scales, and then, it is cascaded towards the small scales and finally is dissipated due to the viscosity. It is also seen that the scaling is the same for all resolutions at sufficiently large scales. By decreasing the length scale (which is equivalent to an increase in wavenumber), this relation changes such that at the intermediate scales, it becomes $w^{-4}$. It should be noted that the very small scales requires higher resolutions, which will be investigated in our future studies. Nevertheless, with the current resolution, the calculated energy spectrum is in a good agreement with the results of previous studies (see for example figure 4 of \cite{universalturbulence}). In passing, it is worthy to state that the curve in \figref{fig:energy} is obtained for the case with small Reynolds number where inertial effects are negligible. We have shown in our unpublished study that the inertia forces notably alter the characteristic of kinetic energy spectrum.

\begin{figure}[hbt!]
	\centering
	\includegraphics[width=5.0in]{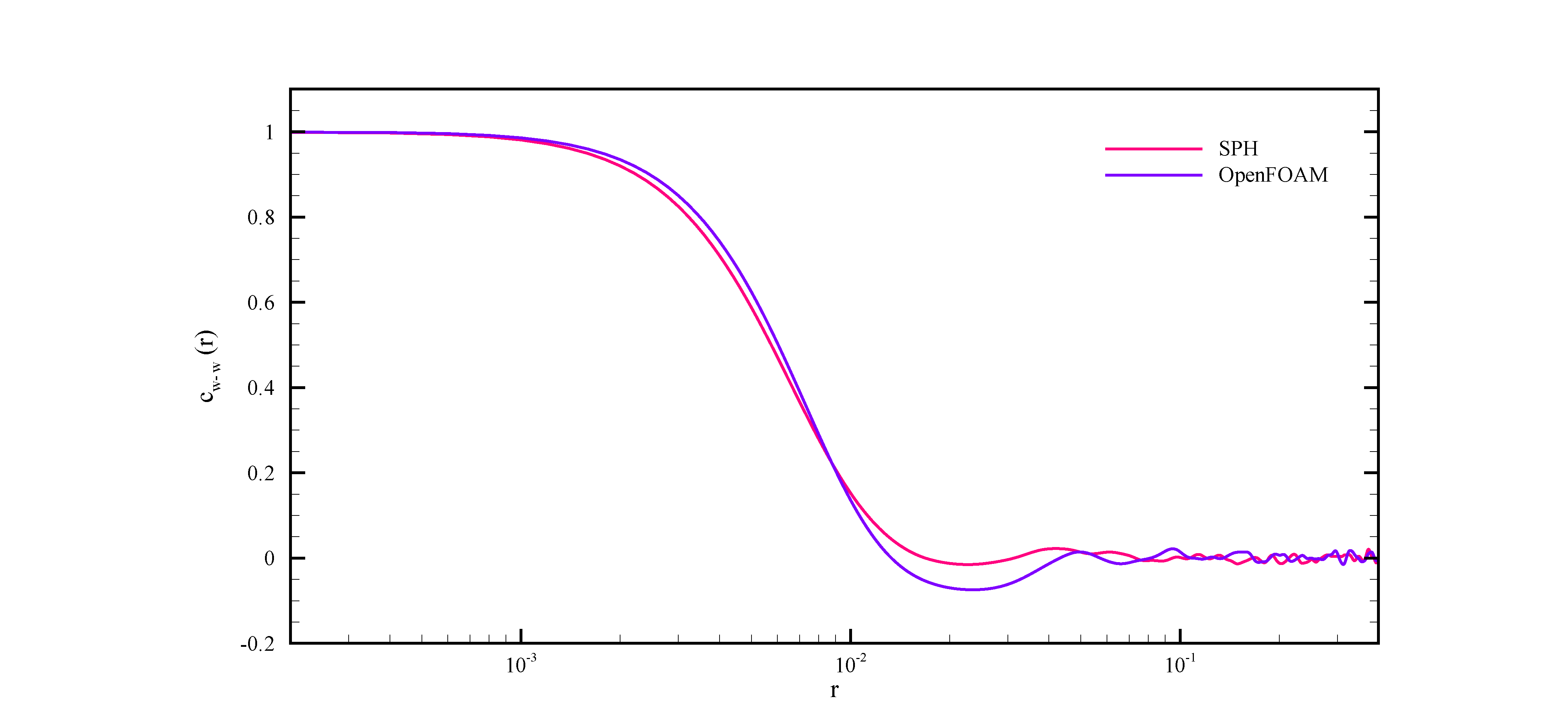}
	\caption{The comparison of vorticity-vorticity correlations calculated by the SPH and OpenFOAM.}
	\label{fig:correlation}
\end{figure}

Referring back to \figref{fig:energy}, one can see that the vortices with various scales exist in the domain. However, it is conventional to select a specific length scale and calculate the important parameters based on it. To study the length scale of the problem, one can use velocity (vorticity)-velocity (vorticity) correlation curve inside the domain.
The normalized correlation function of property $f(r)$ is defined as:

\begin{equation}
	C_{f-f}(r)=\frac{\langle f(r).f(0) \rangle}{f(0)^2},
	\label{eq:corr}
\end{equation}
where $r$ is the spatial position from the point of interest, $r=0$. In the current study $r=0$ is located at the center of the domain and $r$ is selected to vary along the horizontal center-line direction. This function is calculated for the vorticity and is shown in \figref{fig:correlation}. Although similar behavior is observed between our computed correlation and those reported in figure 4 of \cite{nematiccorrelation2}, to further reveal the high fidelity of the proposed SPH model, a similar simulation is also performed using the OpenFOAM package \cite{OpenFOAM} and the results are comparatively provided in \figref{fig:correlation}. As the distance between the point of interest and its all neighbours within the entire flow domain increases, the correlation function tends to become zero, indicating the absence of correlation between two positions spaced far apart, which is also in the agreement with the result of OpenFOAM.

Length scale $(l)$ of active nematics flow can be calculated quantitatively using the correlation function such that $C_{f-f}(l) = 0$ (or in some studies two-tenths of the maximum correlation value
\cite{nematiccorrelation1}) which is the point where the sign of the statistically large distributed vortices is being changed. It is also possible to define the characteristic length scale based on the active fluid properties. As proposed by \cite{scale}, the characteristic length scale can be calculated as $l_Q\approx \sqrt{K/\zeta}$. Also, the characteristic velocity is defined as $v_Q\approx l_Q \zeta /\Gamma$. Consequently, the Reynolds number ($\rnl$) is defined as $\rnl=\rho v_Q l_Q/\mu$. Adopting these scales, the Reynolds number of $\rnl \approx 0.075$ is determined for our simulations.

\begin{figure}[hbt!]
	\centering
	\includegraphics[width=5.0in]{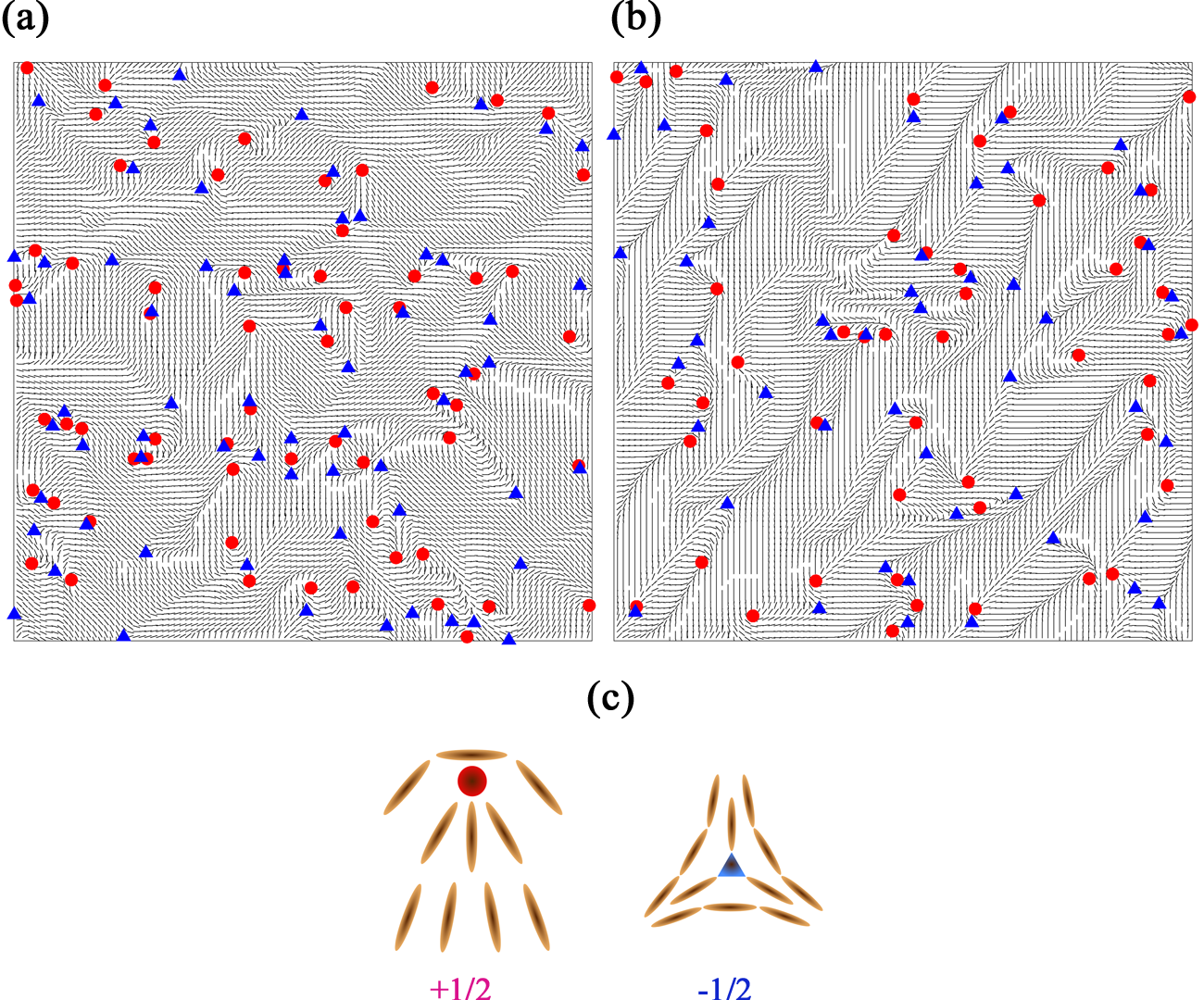}
	\caption{Particle alignment and topological defects of the domain corresponding to $\zeta=0.023$ (a) and $\zeta=-0.023$ (b). A comet-like, $+1/2$, and a trefoil-like, $-1/2$ defects are shown in red filled circles and blue filled triangles respectively. Schematic representation of defects (c).}
	\label{fig:nematicdirector}
\end{figure}

As mentioned earlier, in the study of \cite{universalturbulence}, the defect formation was neglected, which plays an important role in the nematic flow. Thus, herein, we focused on the defect formation in our simulation. The large-scale presentation of the nematic directors is shown in \figref{fig:nematicdirector}-a and -b for positive and negative activities, respectively. As introduced in \secref{sec:int}, topological defects are the important characteristics of active nematics. These defects are the points where there are mismatches between the director of neighboring particles. There are mainly two kinds of defects in the active nematics, positive half and negative half, which are shown by red and blue symbols, respectively in \figref{fig:nematicdirector}. In order to detect these defects in the flow, every single particle is selected and alignments of its surrounding particles are compared with each other. If the surrounding particles alignments are similar to what is shown in \figref{fig:nematicdirector}-c, that particle is labeled as a defect point. The negative half defect has a symmetric structure, so it is balanced inside the flow, whereas the positive half does not have a balanced structure which makes it motile \cite{amin2018}. Because of this motility, when these defects are created in pairs, $+1/2$ pairs move away from each other. Generally speaking, there is a cycle in active nematic flow that describes its behavior; instabilities in the flow field lead to the formation of local wall structures in the nematic director field. Walls are the lines that surround a nematic region and separate these regions from each other, and they are the points where the topological defects are being created by active stress. Due to the gradients in the nematic field around the defects, these defects move and annihilate and restore the nematic order, which again triggers the instability, and this cycle repeats \cite{turb1}.

\begin{figure}[hbt!]
	\centering
	\includegraphics[width=5.0in]{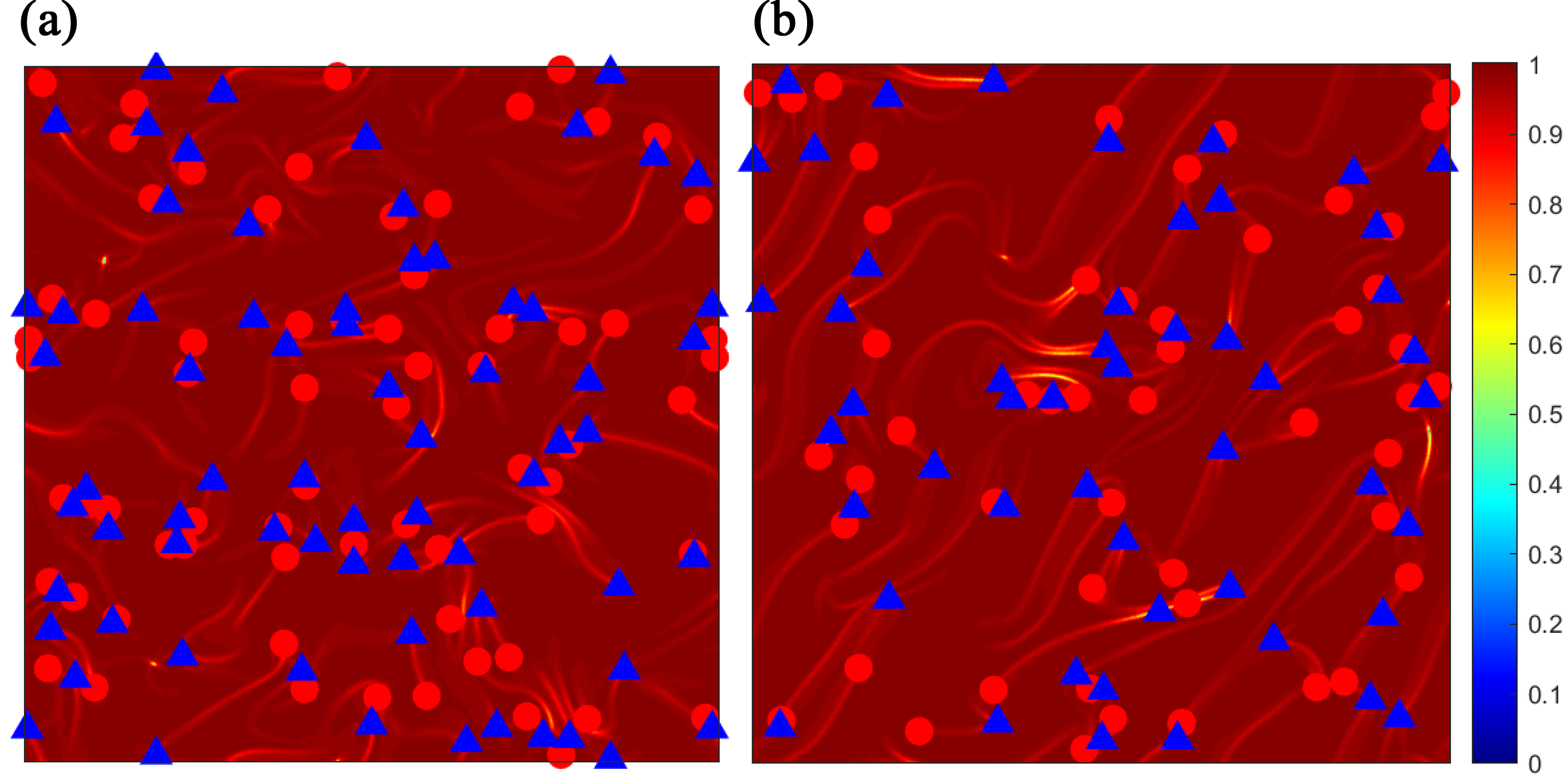}
	\caption{Contour of the nematic order corresponding to $\zeta=0.023$ (a) and $\zeta=-0.023$ (b).}
	\label{fig:nematicorder}
\end{figure}

As stated above, active walls are the places of defect formation, thus it would be beneficial to plot and discuss it in detail. In order to demonstrate the walls inside the domain, it is proper to plot the contour of the nematic order ($q$), as shown in \figref{fig:nematicorder}. This value varies between $0$ and $1$, while the most of domain has the value of $q \approx 1$ which is shown by the dark red color. Along the nematic walls, on the other hand, $q$ gets a smaller value, which is colored by light red in \figref{fig:nematicorder}. It is important to note that, as mentioned earlier, all the defects are located on the wall where $q \neq 1$.

Since the activity induces energy into the nematic flow and influences its characteristics, it is important to evaluate its effect on the flow.
\begin{figure}[hbt!]
	\centering
	\includegraphics[width=5.0in]{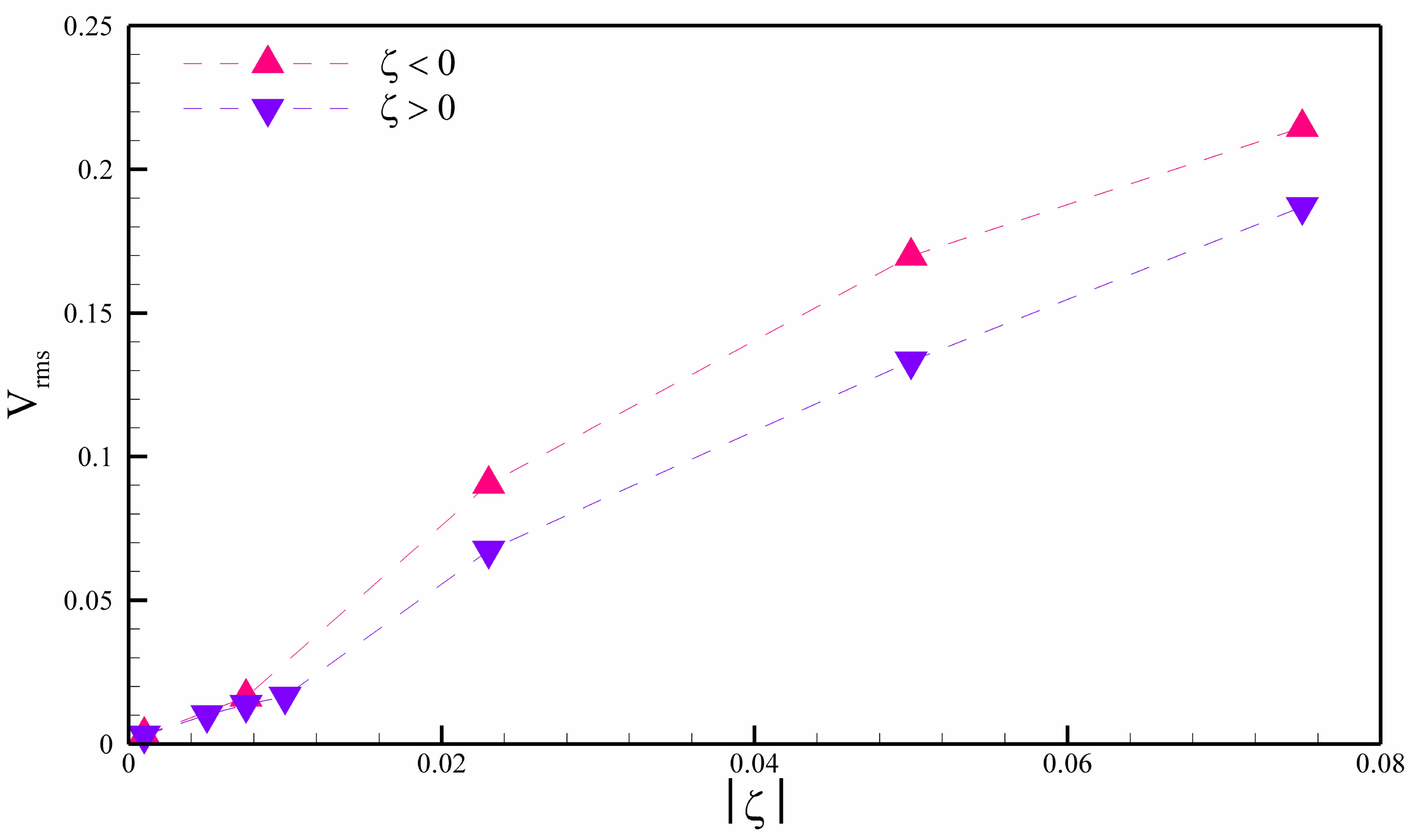}
	\caption{Effect of activity on the root mean square of velocity for extensile and contractile suspensions.}
	\label{fig:vrms}
\end{figure}
To define the characteristic velocity in turbulent-like flow, it is common to use the root mean square of velocity ($V_{rms}$), which, as an averaging term, includes both positive and negative velocity fluctuations. This property is calculated under the effect of various amounts of activity and demonstrated in \figref{fig:vrms} for extensile ($\zeta>0$) and contractile ($\zeta<0$) suspensions. As it is seen in this figure, both extensile and contractile nematics behave similarly, which means that the increase in the absolute value of activity elevates the value of $V_{rms}$. It should be noted that the dependency of $V_{rms}$ on the activity is stronger for contractile than for the extensile, hence $\zeta<0$ stays above the $\zeta>0$ curve. Moreover, it is seen that these graphs increase monotonically with respect to activity. 

\begin{figure}[hbt!]
	\centering
	\includegraphics[width=2.0in]{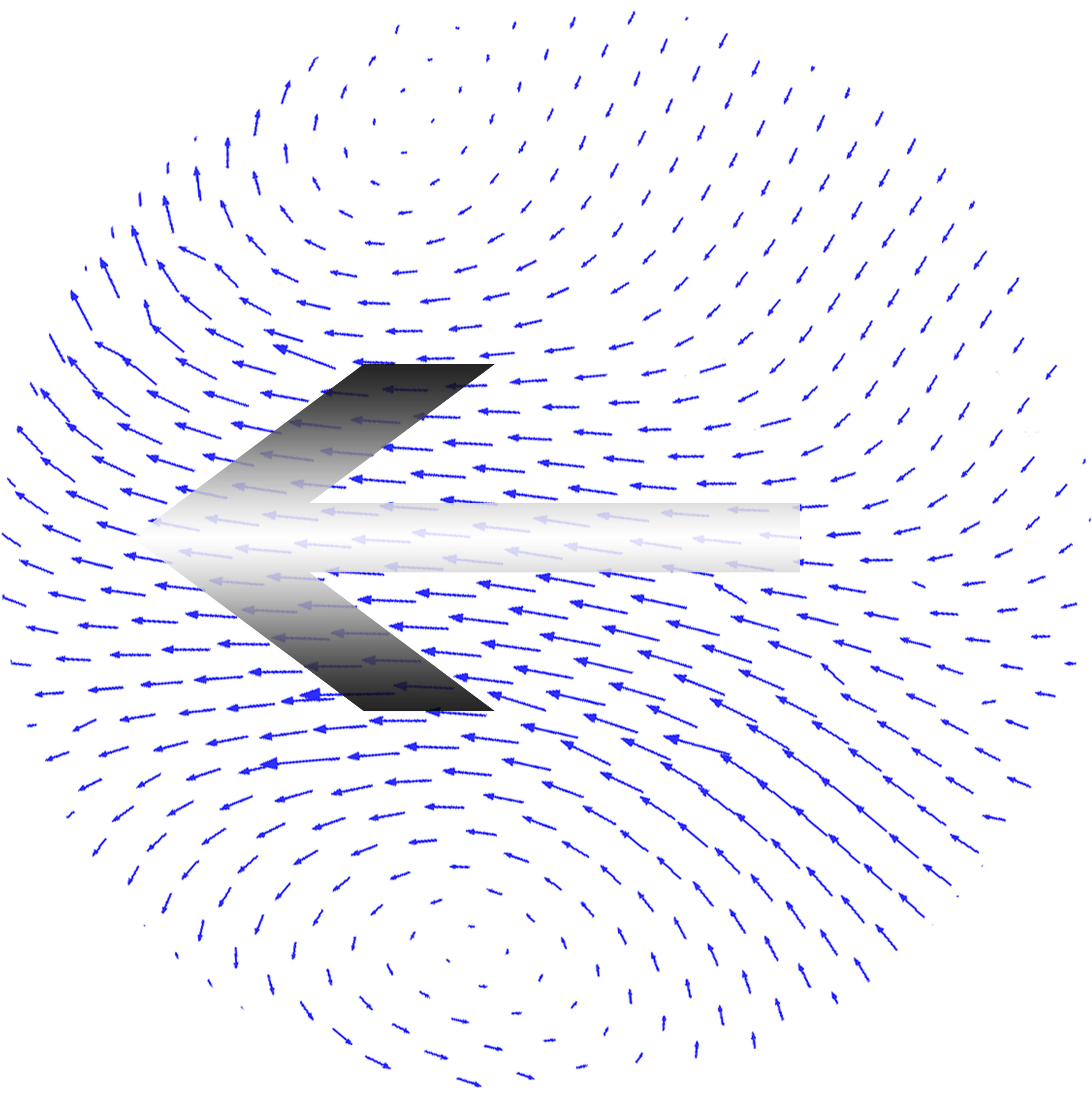}
	\caption{Average $+1/2$ defect flow for $\zeta = 0.05$.}
	\label{fig:averagedefectflow}
\end{figure}

The dependency of $V_{rms}$ on the activity is directly related to the velocity jets induced in nematics by defects in the nematic flow. The average flow induced by the defects is calculated for $+1/2$ defects and shown in \figref{fig:averagedefectflow}. Typically $+1/2$ defects are associated with a vortex dipole as shown in this figure. This defect is self-propelled in the direction shown by the arrow. The flow pattern presented in \figref{fig:averagedefectflow} is consistent with that obtained by an analytical solution using Green's function~\cite{Giomi2015}.

\begin{figure}[hbt!]
	\centering
	\includegraphics[width=4.0in]{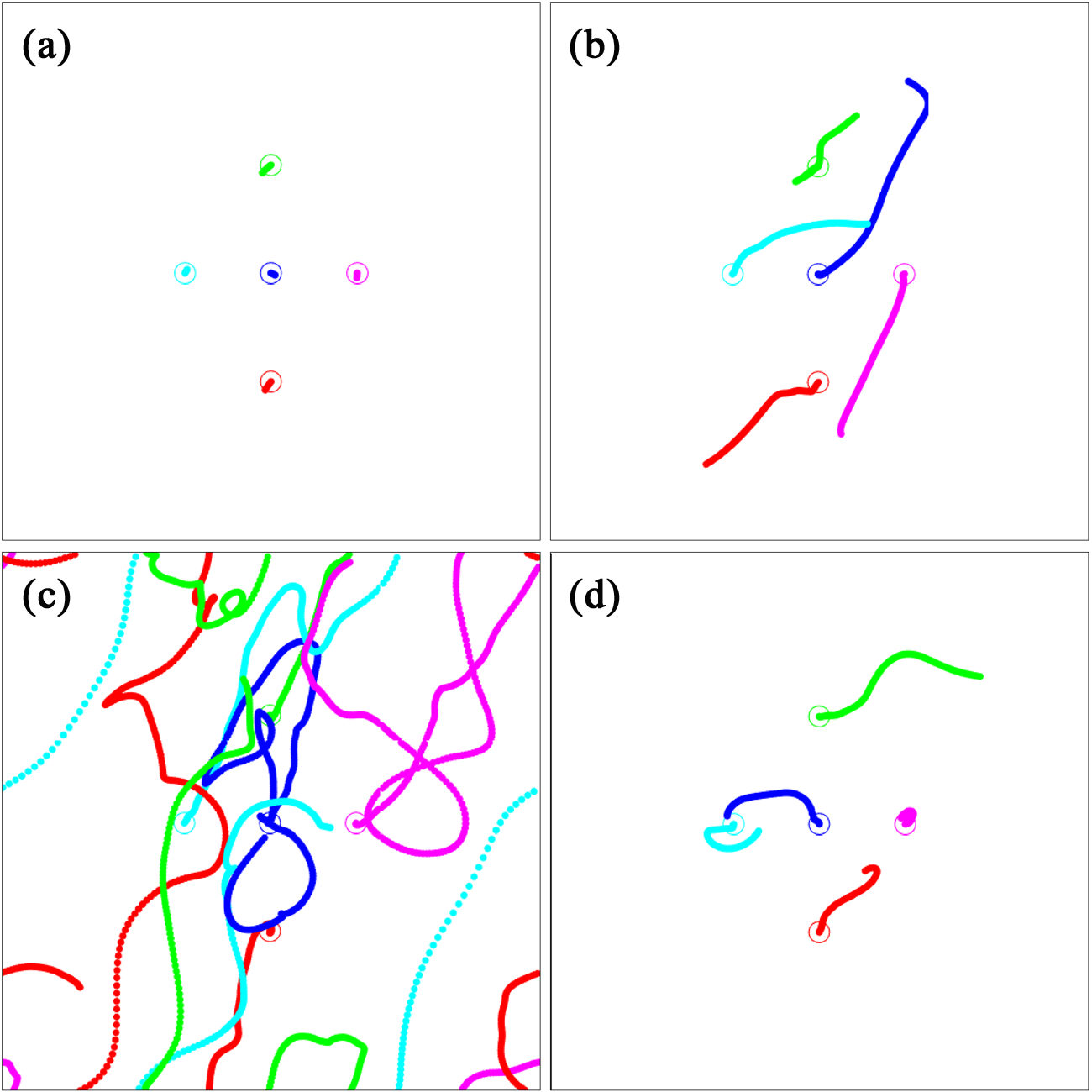}
	\caption{Nematic particles pathline at the same physical time for $\zeta = -0.001$ (a), $\zeta = -0.023$ (b), $\zeta = -0.25$ (c), and $\zeta = 0.023$ (d).}
	\label{fig:fig4-tracking}
\end{figure}

As mentioned in \secref{sec:int}, the important advantages of SPH over other numerical methods is its Lagrangian nature. With this feature, one can track every single nematic particle inside the flow and evaluate the variation of its properties during the simulation. To exploit this feature, we took five nematic particles from different initial positions and track their trajectories during the simulation. Pathlines of these particles are shown in \figref{fig:fig4-tracking} for four different activities, $\zeta = -0.001, \pm0.023$, and $-0.25$. Initial positions are encircled and each point in pathlines represents a particle position in a specific time step. As expected, by increasing the absolute value of activity ($\abs{\zeta}$), lengths of pathlines increase. This happens due to the high energy injection at higher activities. Nematic particles consume this energy to move faster and consequently further. At the smaller values of activity, however, viscous term dominates the flow and dissipates the input energy. As a result, the pathlines of the nematic particles are shorter for the smaller values of activity. Interestingly, the pathline for the $\zeta = -0.023$ is longer than of $\zeta = 0.023$ which was previously observed in \figref{fig:vrms} which explains the difference between the $V_{rms}$ curves for extensile and contractile suspensions. This shows that the contractile ($\zeta<0$) suspension converts more input energy into kinetic energy which leads to a larger velocity and displacement.

\begin{figure}[hbt!]
	\centering
	\includegraphics[width=5.0in]{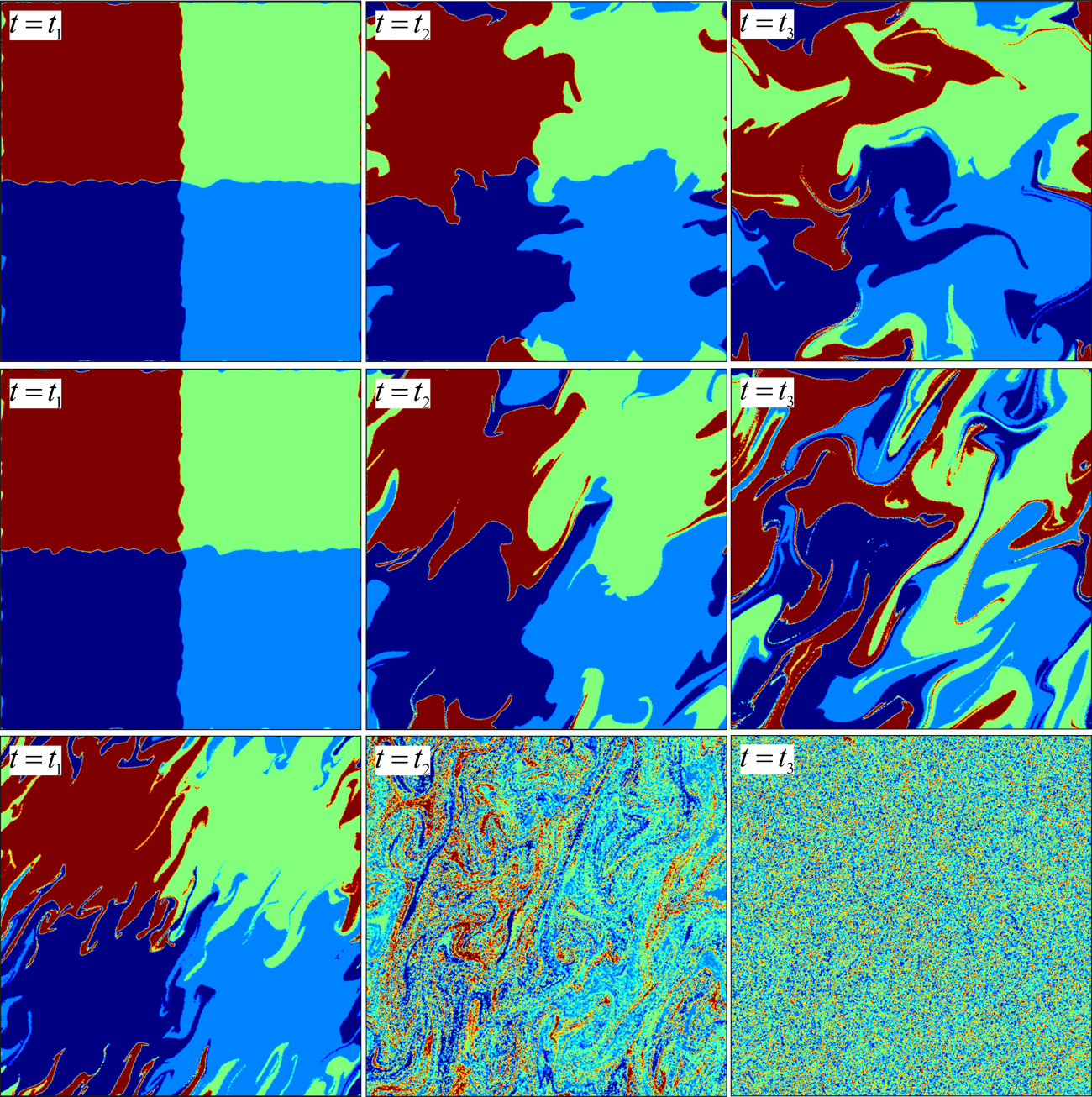}
	\caption{Comparison of mixing for $\zeta = 0.023$ (first row), $\zeta = -0.023$ (second row), and $\zeta = -0.25$ (last row). Each column corresponds to the identical physical time with $t_1<t_2<t_3$.}
	\label{fig:fig6-mixingcomp}
\end{figure}

In order to track the large groups of particles, we divided the domain into four regions and colored the particles inside of each region to distinguish the particles during the simulation. This leads to the mixing of the particles as shown in \figref{fig:fig6-mixingcomp}. It should be emphasized that the particles inside the whole domain have identical nematic and hydrodynamical properties and their color only represents their initial position. While advancing in time, the four regions start to blend with each other, and by increasing time adequately, a uniform distribution of particles from all regions is obtained, which demonstrates the uniform mixing of the nematic particles. It can be inferred that the particles are free to move in any direction, but their mutual interactions dictate which direction they move in. Moreover, the dynamics are in fact deterministic, the seemingly-randomness comes from the fact that the system is chaotic.
\figref{fig:fig6-mixingcomp} also demonstrates the effect of activity strength on the mixing behavior of active nematics. $t_1=45<t_2=252<t_3=500$ Sec. are the real physical times at which snapshots of the particle positions are taken. As expected, the case with the higher value of the activity reaches complete mixing faster whereas those with the lower values of activity need a comparatively longer duration. The comparison of the first and the second row of \figref{fig:fig6-mixingcomp} indicates the high capability of contractile nematics for the mixing. Thus, for the mixing purpose, contractile nematics perform better than the extensile one and the operation speed can be hastened by increasing the activity.

\begin{figure}[hbt!]
	\centering
	\includegraphics[width=5.0in]{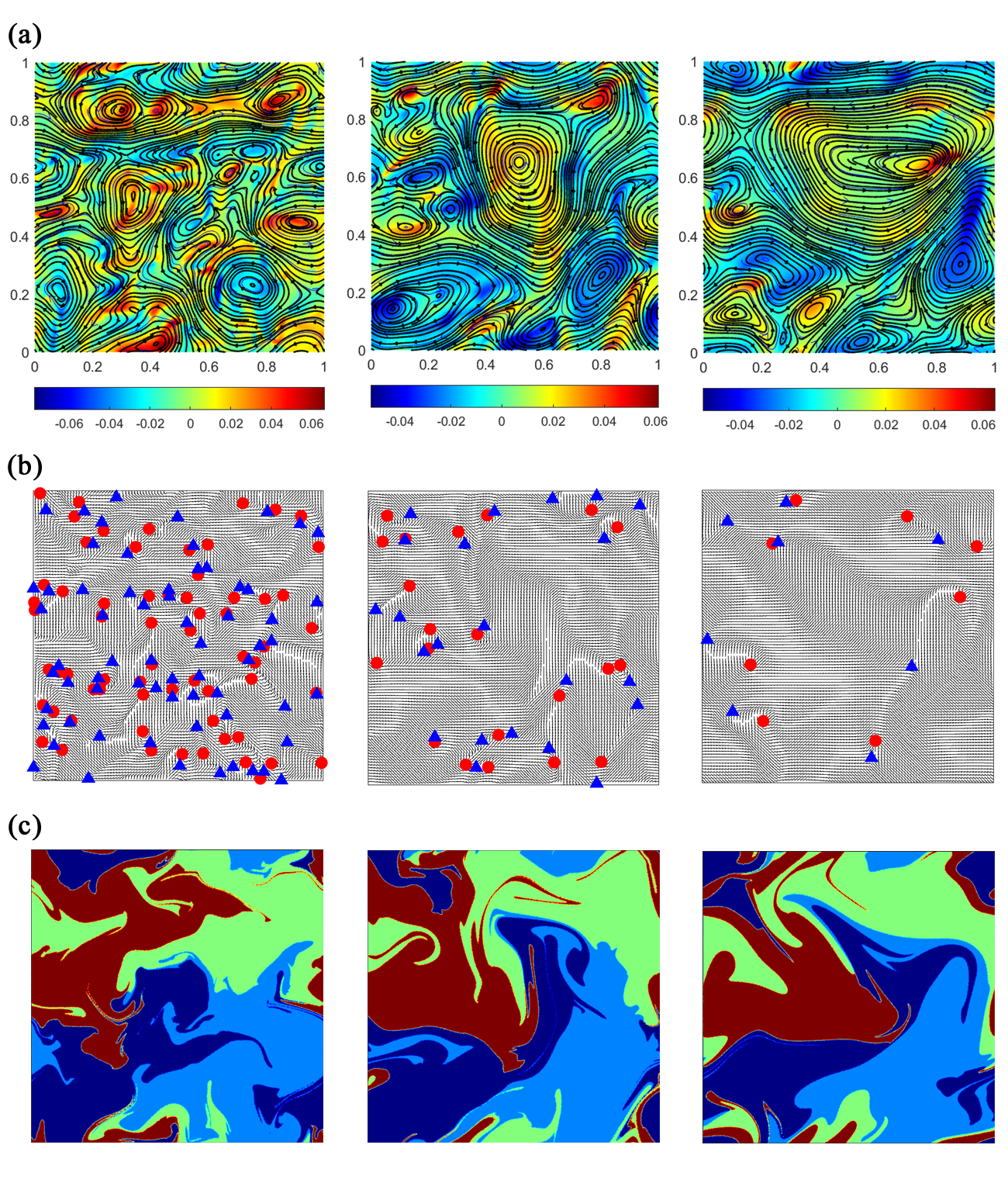}
	\caption{Effect of elastic constant on the contours of vorticity and streamlines (a), topological defects and nematic ordering (b), and mixing (c). First, second and third columns represent $K = 0.02$, $K = 0.05$, and $K = 0.1$, respectively.}
	\label{fig:keffect}
\end{figure}

Thus far, we have focused on the effect of activity on the active nematic flow characteristics. As described in \secref{sec:gov}, in the formulation of the nematodynamics, elastic constant, $K$ was introduced, which plays an important role in the molecular field tensor, $\veb{H}$. \figref{fig:keffect}-a demonstrates the variation of the flow field with $K$. As can be seen from this figure, the increase in the value of $K$ augments the vortex sizes, which is compatible with the relation $\sqrt{K/\zeta}$ describing the length scale of the nematic flow. 
Physically, this trend can be investigated considering the energy consumption in active nematics. The injected energy by activity is consumed for the creation of topological defects but at the expense of free energy. Consequently, as shown in \figref{fig:keffect}-b, defect number strongly depends on the value of $K$ and decreases with an increase in $K$. Moreover, as discussed earlier, there is a coupling between defects and fluid instabilities by creation and annihilation cycle of topological defects. These instabilities, on the other hand, are the reason of the vortex creation. By decreasing the defect number, instabilities become weaker and a few vortices are created in the flow with the larger size hence covering the whole domain.
Similarly, lower values of $K$ result in higher instability in which the flow is much more chaotic. This behavior is demonstrated in \figref{fig:keffect}-c which shows the snapshots of the flow mixing for different values of $K$. It should be noted that all of the snapshots are taken at the same physical time, $t=500$. Interfaces between different colors exhibit the instability of the flow. As shown in this figure, the interface at $K = 0.1$ is smoother in comparison with the flow at $K = 0.02$, which justifies the inverse effect of $K$ on the instability of the flow. Thus, to increase the mixing efficiency, it is suggested to use the smaller value of $K$.
%Physically, this trend can be investigated considering the energy consumption in active nematics. The injected energy by activity is consumed for the creation of topological defects and breaking down the vortices to smaller scales but at the expense of free energy. On the other hand, the presence of $K$ in Eq. \eqref{eq:freeenergy} determines the cost of the free energy minimization. Consequently, $K$ confronts the creation of low-scale vortices as well as topological defects in the flow. This is the reason for the creation of large-scale vortices at the high $K$ value in \figref{fig:keffect}-a.
%Similarly, as shown in \figref{fig:keffect}-b, defect number strongly depends on the value of $K$ and decreases with an increase in $K$. As mentioned previously, active energy results in the formation of topological defects that can move inside the fluid and creates instability. Consequently, higher magnitude of $K$ results in smaller instability while for the lower values of $K$, the flow is much more chaotic. This behavior is demonstrated in \figref{fig:keffect}-c which shows the snapshots of the flow mixing for different values of $K$. It should be noted that all of the snapshots are taken at the same physical time. Interfaces between different colors exhibit the instability of the flow. As shown in this figure, the interface at $K = 0.1$ is smoother in comparison with the flow at $K = 0.02$, which justifies the inverse effect of $K$ on the instability of the flow. Thus, to increase the mixing efficiency, it is suggested to use the smaller value of $K$.

\section{\label{sec:conc}Conclusions}
In this study, the weakly compressible smoothed particle hydrodynamics method is used to simulate the active nematic fluid. A parallel, object-oriented computer code is developed using the CUDA C++ platform in order to accelerate the simulations. Ghost periodic boundary condition is introduced and applied to all boundaries to imposed the periodic boundary condition.
Since the nematic fluid is composed of nematic particles, collective movement of SPH particles is perfectly mimicked the movement of the nematic particles. Fluid flow characteristics, including vortex structures and streamlines, are exhibited, and the turbulent-like behavior of the nematic fluid is presented qualitatively. To scrutinize the turbulent characteristics, the relation between the kinetic energy and wavenumber is evaluated, and results are observed to agree with the universal relations for different ranges of length scales. Vorticity- vorticity correlation is evaluated, and the characteristic length scale is defined based on it. Nematic orders and director is calculated and used to detect the topological defects. Positive half and negative half defects are discussed in detail. Results are presented for the effects of two important parameters, activity and elastic constant. The effect of activity on the velocity root mean square is evaluated, and it is seen that by increasing the absolute value of the activity, the velocity root mean square increases as well, while its effect is strong for negative activities. To exploit the SPH capabilities, pathlines and mixing of nematic particles are described qualitatively. It was shown that the length of pathlines is proportional to the activity. The effect of the elastic constant is also calculated, and it is shown that higher values of elastic constant exhibit larger vortex length scales and smaller defect numbers, and inferior mixing. It was shown that the activity and elastic constant behave oppositely in the creation of chaotic flow, which initiates from the energy minimization.

%Bibliography
%\bibliographystyle{unsrt}  
%\bibliography{cas-refs}  

\end{document}